\RequirePackage{fix-cm}
\documentclass[smallextended]{svjour3}       
\smartqed  
\usepackage{graphicx}
\usepackage{subfigure}
\usepackage{amsmath}
\makeatletter
\renewcommand{\@thesubfigure}{\hskip\subfiglabelskip}
\makeatother
\begin{document}

\title{Cooperation on the monte carlo rule: Prison's dilemma game on the grid
}


\author{Jiadong Wu         \and
        Chengye Zhao 
}


\institute{Jiadong Wu \at
              China Jiliang University\\
              \email{rxywjd1901@163.com}           
           \and
           Chengye Zhao \at
              China Jiliang University\\
              \email{cyzhao@cjlu.edu.cn(corresponding author)}
} 

\date{Received: date / Accepted: date}

\maketitle

\begin{abstract}
In this paper, we investigate the prison's dilemma game with monte carlo rule in the view of the idea of the classic Monte Carlo method on the grid.
    Monte carlo rule is an organic combination of the current dynamic rules of individual policy adjustment, which not only makes full use of information but also reflects the individual's bounded rational behavior and the ambivalence between the pursuit of high returns and high risks. In addition, it also reflects the individual's behavioral execution preferences.
    The implementation of monte carlo rule brings an extremely good result, higher cooperation level and stronger robustness are achieved by comparing with the unconditional imitation rule, replicator dynamics rule and fermi rule.
    When analyse the equilibrium density of cooperators as a function of the temptation to defect, it appears a smooth transition between the mixed state of coexistence of cooperators and defectors and the pure state of defectors when enhancing the temptation, which can be perfectly characterized by the trigonometric behavior instead of the power-law behavior discovered in the pioneer's work.
    When discuss the relationship between the temptation to defect and the average returns of cooperators and defectors, it is found that cooperators' average returns is almost a constant throughout the whole temptation parameter ranges while defectors' decreases as the growth of temptation.
    Additionally, the insensitivity of cooperation level to the initial density of cooperators and the sensitivity to the social population have been both demonstrated.
\keywords{Cooperation \and Prison's dilemma \and Monte carlo rule \and Robustness}
\end{abstract}

\section{Introduction}
The cooperation among selfish individuals [1] such as Kin cooperation [2], Mutually cooperation and Reputation-seeking cooperation [3], which are contrary to natural selection, are widely discovered in human society [4-6], having already become one of the challenges of evolutionary game theory [7-9].
\par
This phenomenon can be characterized by the prison's dilemma game [10], in which the individuals adopt one of two strategies: C (cooperation) or D (defection). A selfish individual would select D as his strategy for the sake of higher returns. Nevertheless, if both sides choose D simultaneously, each of them will get less returns than those acquired for mutual cooperation.
\par
Over the years, scholars often focus exclusively on the promotion of cooperation on different spatial structures [11-13]. However, there are not many researches on the dynamic rules of game individual policy adjustment.
Different update rules often lead to different results. For example, Sysi-Aho et al. [14] have modified a more rational updating rule on the basis of the research of Harut and Doebeli [15], believing that the individuals have original intelligence in the form of local decision-making rule deciding their strategies. In this rule, individuals suppose their neighbors' strategies retain unchanged and aim at choosing a strategy to maximize their instant returns. This rule results in the density of cooperator at equilibrium which differ tremendously from those resulting from the replicator dynamics rule in the literature [15], and the cooperation can persist throughout the whole temptation parameter ranges.
Li et al. [16] adopt the unconditional imitation rule used in the Nowak and May's work [17] to revise the regulation of replication dynamics owing to its briefness and the ability of according with the psychology of most individuals. It is discovered that in some parameter ranges, the performance of inhibiting cooperative behaviors originally turns to promoting.
Xia et al. [18] compare the effect on cooperation under unconditional imitation rule, replicator dynamics rule and Moran process [19] respectively, discovering that Moran process promotes cooperation much more than the others.
Those facts affirm the status of update rules in the evolution of cooperative behavior.
\par
In the existing research, the dynamic rules of game individual strategy adjustment are mainly unconditional imitation rule, replicator dynamics rule and fermi rule [20]. However, replicator dynamics and fermi rule can not make full use of the game information while unconditional imitation rule can not tolerate individuals' irrational behavior. An excellent update rule that simulates individual policy updates more perfectly needs to be discovered.
\par
In this paper we propose monte carlo rule, which is an organic combination of the current dynamic rules of individual policy adjustment, not only making full use of information but also reflecting the individual's bounded rational behavior and the ambivalence between the pursuit of high interest and high risk [21,22]. In addition, it also reflects the individual's behavioral execution preferences [23,24].
We analyse the effect on cooperative level under monte carlo rule and verify its robustness. Further, we analyse the equilibrium density of cooperator as a function of the temptation to defect and use trigonometric curves to characterize it. It is confirmed that the trigonometric fitting effect is better than the power-law fitting in the pioneer's work [20].
We also investigate the relationship between temptation to defect and the average returns of cooperators and defectors. Additionally, the insensitivity of cooperation level to the initial density of cooperators and
the sensitivity to the social population have been both demonstrated by numerical simulation.
\section{Prison's dilemma game model with monte carlo rule}
The prison's dilemma represents a class of game models that its Nash equilibrium only falls on the non-cooperation. In this game model, each individual can adopt one of two strategies C (cooperation) and D (defection). The returns depend on the strategy of both sides. Enormous temptation forces rational individuals to defect. However, if both sides choose the strategy of D simultaneously, each of them will get less returns than those acquired for mutual cooperation.
\par
Game individuals are located on the nodes of the grid, and the edges indicate the connections between one individual and another. Along the footsteps of Nowak et al. [25], the game returns can be simplified as the matrix below:
\par
\begin{table}[h]
\centering
\begin{tabular}{clc}
 & \vline  \ \ \ Cooperate & Defect \\ \hline
Cooperate&\vline \ \ \ \ \ \ \ \ \ $1$ & $0$ \\
Defect&\vline\ \ \ \ \ \ \ \ \  $b$ & $0$ \\
\end{tabular}
\end{table}
Where parameter $b$ characterizes the interests of the temptation to defect ($1<b\le2$).
The larger the value of $b$, the greater the temptation of defection to the individuals.
\par
Individuals gain returns by playing prisoner's dilemma game with their nearest neighbors. During the evolutionary process, each individual adopts one of neighbors' strategies whose returns is more than or equal to himself's in the way of roulette, or just insists the original strategy:
\begin{equation}
P(s_i \leftarrow s_j)=\frac{U_j}{U_0+\sum\limits_{k=1}\limits^{d}\omega (i,k)U_k}\ \
,\ \ j=0,1,2,...,d
\end{equation}
Where $U_0$ represents individual's own returns and $U_1,...,U_d$ represent the returns of the nearest neighbors respectively, $d$ expressing the dimension of the rule network (that is, $d$=4 in this paper) and $P(s_i \leftarrow s_j)$ showing the probability that individual $i$ would imitate $j$. $\omega (i,k)$ is a characteristic function:
\begin{equation}\omega (i,k)=
\begin{cases}
0 \ \ &U_i>U_k,
\\
1 \ \ &U_i\le U_k.
\end{cases}
\end{equation}
Where $U_i$ represents individual $i$'s returns and $U_k$ ($k=1,2,...,d$) represents the returns of individual $k$ who is one of $i$'s neighbors.
\section{Results and Discussions on Simulation Experiment}

For the sake of investigating the influence of spatial structure to cooperative behavior, we apply classical Mean Field Theory [26] which is insensitive to the topology, to preliminary predict the density of cooperators characterized by $\rho$.
Under this circumstance, the average returns of cooperators can be expressed as:
\begin{equation}
U_C=d\rho
\end{equation}
Where $d$ characterizes the dimension of the rule network. In the same way, the average returns of defectors can be expressed as:
\begin{equation}
U_D=bd\rho\
\end{equation}
Where the $b$ characterizes the temptation to defect.
\par
Following the monte carlo rule, we have obtained the differential equation of $\rho$:
\begin{equation}
\begin{aligned}
\dfrac{\partial\rho}{\partial t} &=(1-\rho)W(D\leftarrow C)-\rho W(C\leftarrow D)\\
&=-\rho(1-\rho)\dfrac{d \cdot U_D}{d(1-\rho)U_D+(1+d\rho)U_C}\\
&=-\rho(1-\rho)\dfrac{1}{1-\rho+\dfrac{\rho}{b}+\dfrac{1}{bd}}
\end{aligned}
\end{equation}
Where $W(D\leftarrow C)$ indicates the probability that a defector transforms into a cooperator and $W(C\leftarrow D)$ indicates the probability that a cooperator transforms into a defector. $t$ is the game round.
\par
On the basis of the differential equation, it turns out that as the game progresses, the density of the cooperator ($\rho$) decreases monotonically until it approaches 0, that is, all the cooperators would go extinct (dotted line in Fig. 1).
However, in the space rule networks, cooperators can survive in the form of clusters where they can get support from peers (solid line in Fig. 1), reaffirming the positive role of spatial structure in cooperative behavior[16].
\begin{figure*}[!h]
\centering
\includegraphics[width=0.75\textwidth]{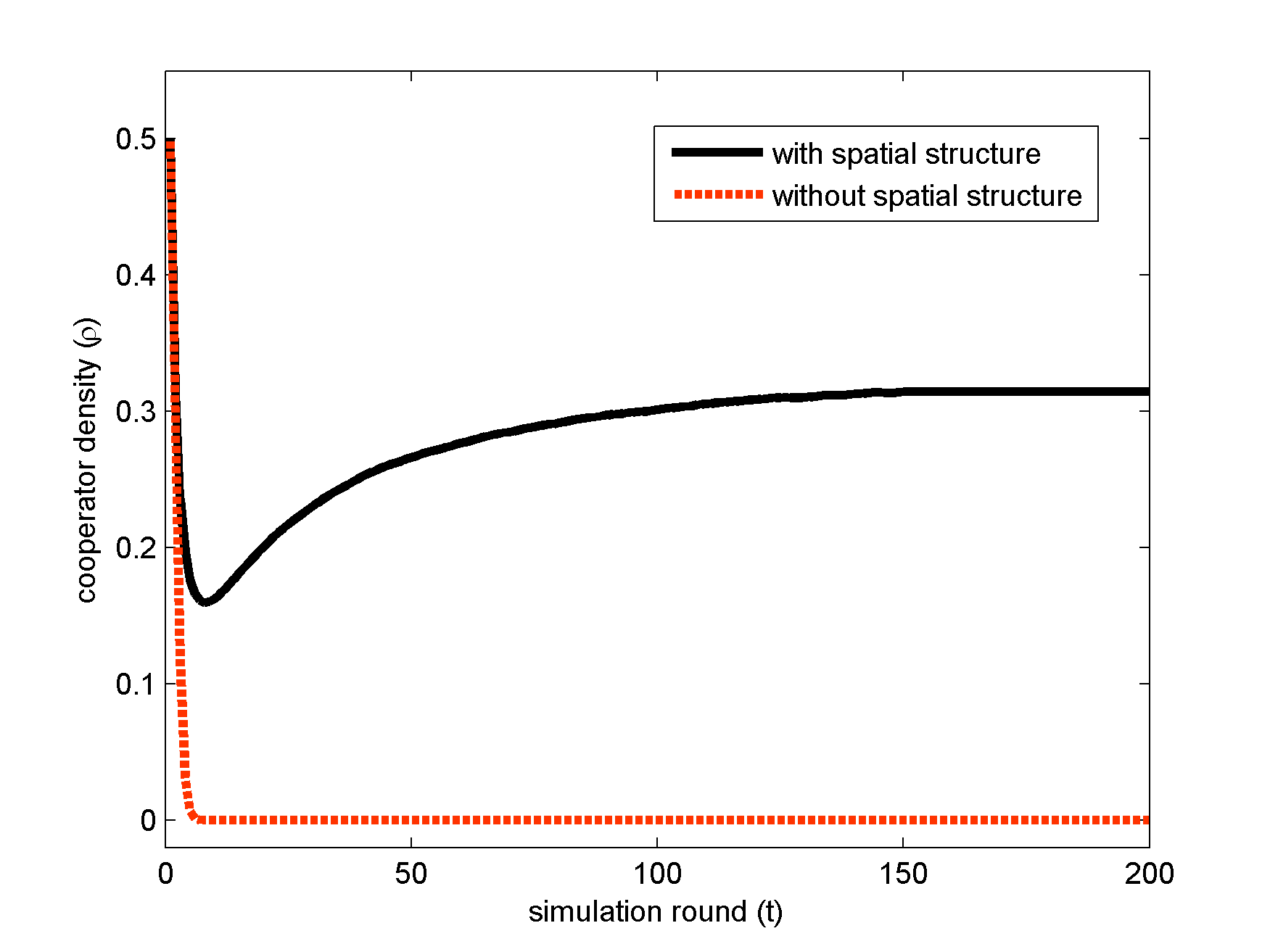}
\caption{Numerical simulation of cooperator density under monte carlo rule (solid line) at b=1.10. The dotted line represents the well-mixed case obtained by Mean Field Theory in the same parameter environment. Those data are simulated on grid of 100x100 where 50\% cooperators as well as 50\% defectors are randomly distributed at the beginning, and 1000 simulations are averaged in each case.}
\label{fig:1}
\end{figure*}
\begin{table}[htb]
\caption{A brief comparison of the four update rules in this paper.}
\label{tab:1}       
\begin{tabular}{lll}
\hline\noalign{\smallskip}
Update strategy & Information utilization & Irrationality \\
\noalign{\smallskip}\hline\noalign{\smallskip}
unconditional imitation & full & not exists  \\
replicator dynamics & not full & exists \\
fermi & not full & exists \\
monte carlo& full & exists \\
\noalign{\smallskip}\hline
\end{tabular}
\end{table}
\par
Similar to the unconditional imitation rule, monte carlo rule makes full use of the the game information when compared with fermi rule and replicator dynamics rule (Table 1). That is, during the policy update phase, individuals would collect game information from all the neighbors to determine the most satisfactory strategy in the next game round, instead of just randomly selecting a neighbor to decide whether to imitate or not. This behavior reflects the rigor of individuals. Also, individual's psychology of pursuing return growth is reflected vividly in the monte carlo rule. Therefore, the game system under the rule can support the germination of cooperation to a large extent. Figure 2 shows this fact: again the social population $N$=10000 and $b$=1.10, yet there is only one very small cooperative group in the middle of the network at t=0, cooperative behavior can still spread promptly.
The middlemost subgraph in Fig. 2 shows the track of function $\rho(t)$, and other subgraphs display the distribution of cooperators and defectors at several important simulate moments. At the beginning, the small cooperative group spreads the cooperative behavior in the form of clusters with irregular shapes. At the edge of the clusters, the cooperators resist the temptation of the outside world through the support of peers in the clusters. Owing to the protection of marginal cooperators, the individuals within the clusters are undoubtedly the loyal defenders of cooperation. This model of mutual support enables cooperators to survive in the society.
As the game progresses, the rate of propagation continues to increase, reaching the peak at around t=500. Then the rate is slowly reduced to 0, achieving dynamic balance at around t=800. It is clearly that driven by the idea of pursuing progress and full game information, the cooperative behavior quickly spreads throughout the whole network.
\begin{figure*}[!h]
\centering
\subfigure[t=0]{
  \includegraphics[width=1.4in]{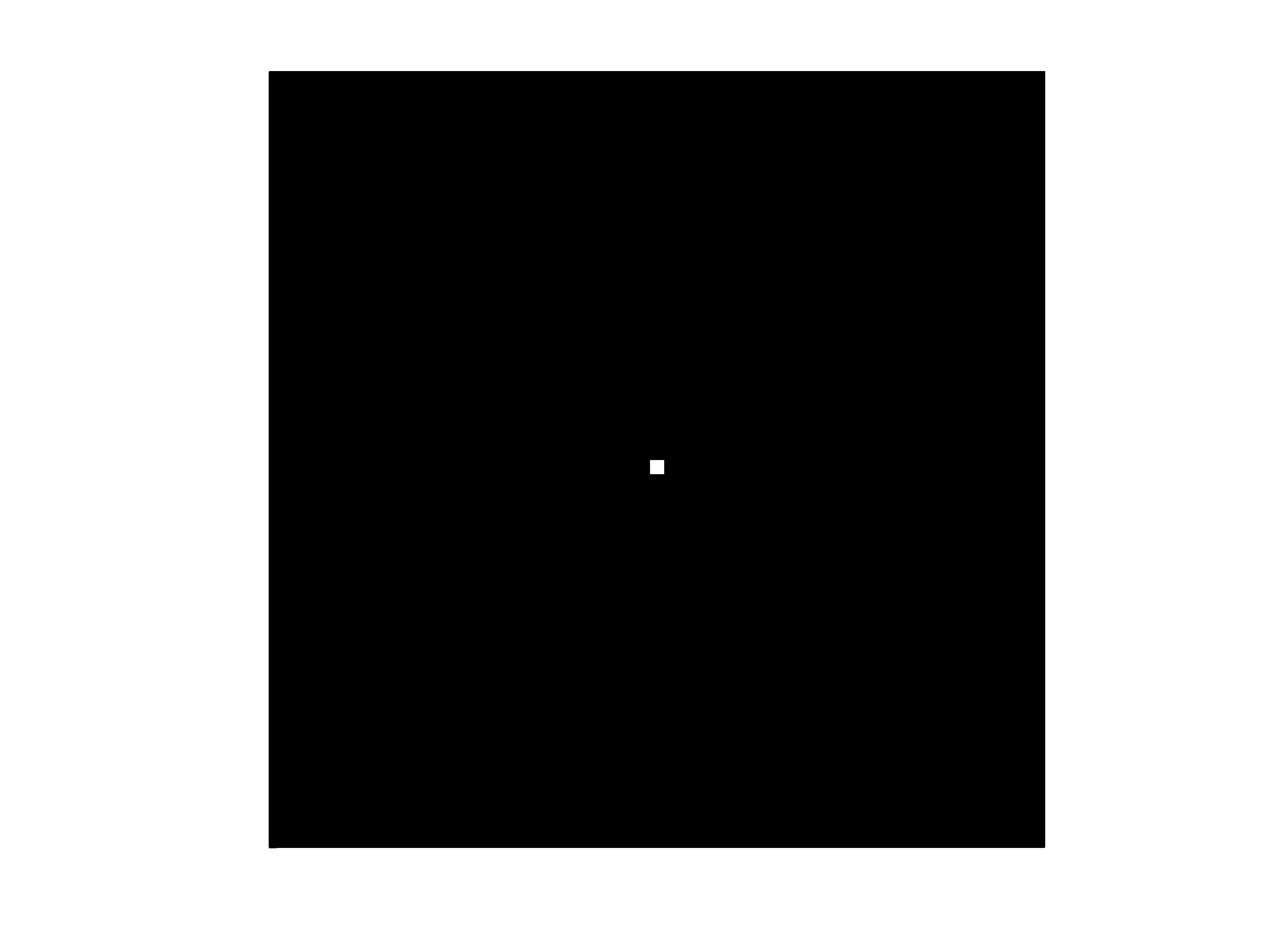}}
\centering
\hfill
\subfigure[t=20]{
  \includegraphics[width=1.4in]{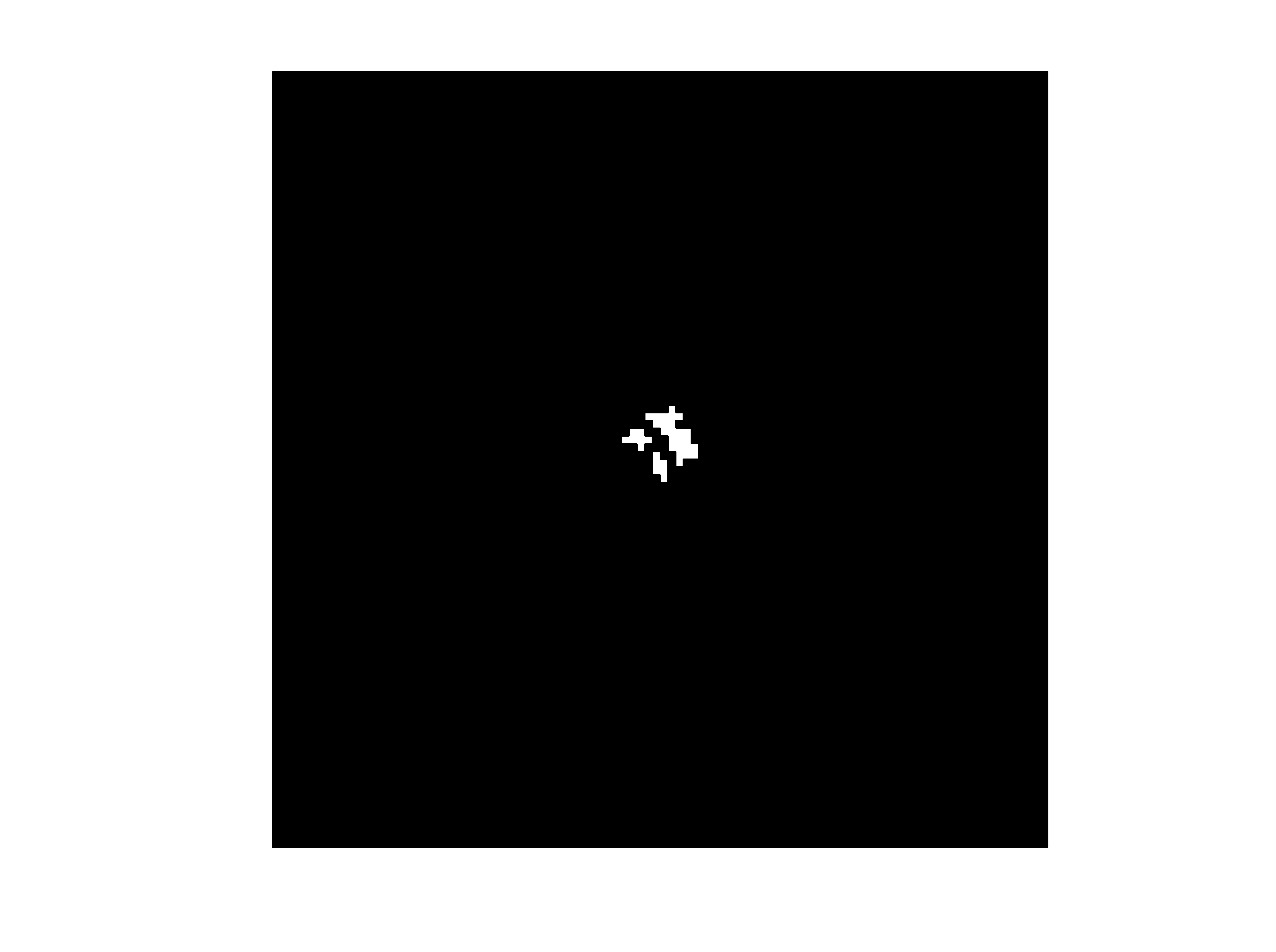}}
\centering
\hfill
\subfigure[t=100]{
  \includegraphics[width=1.4in]{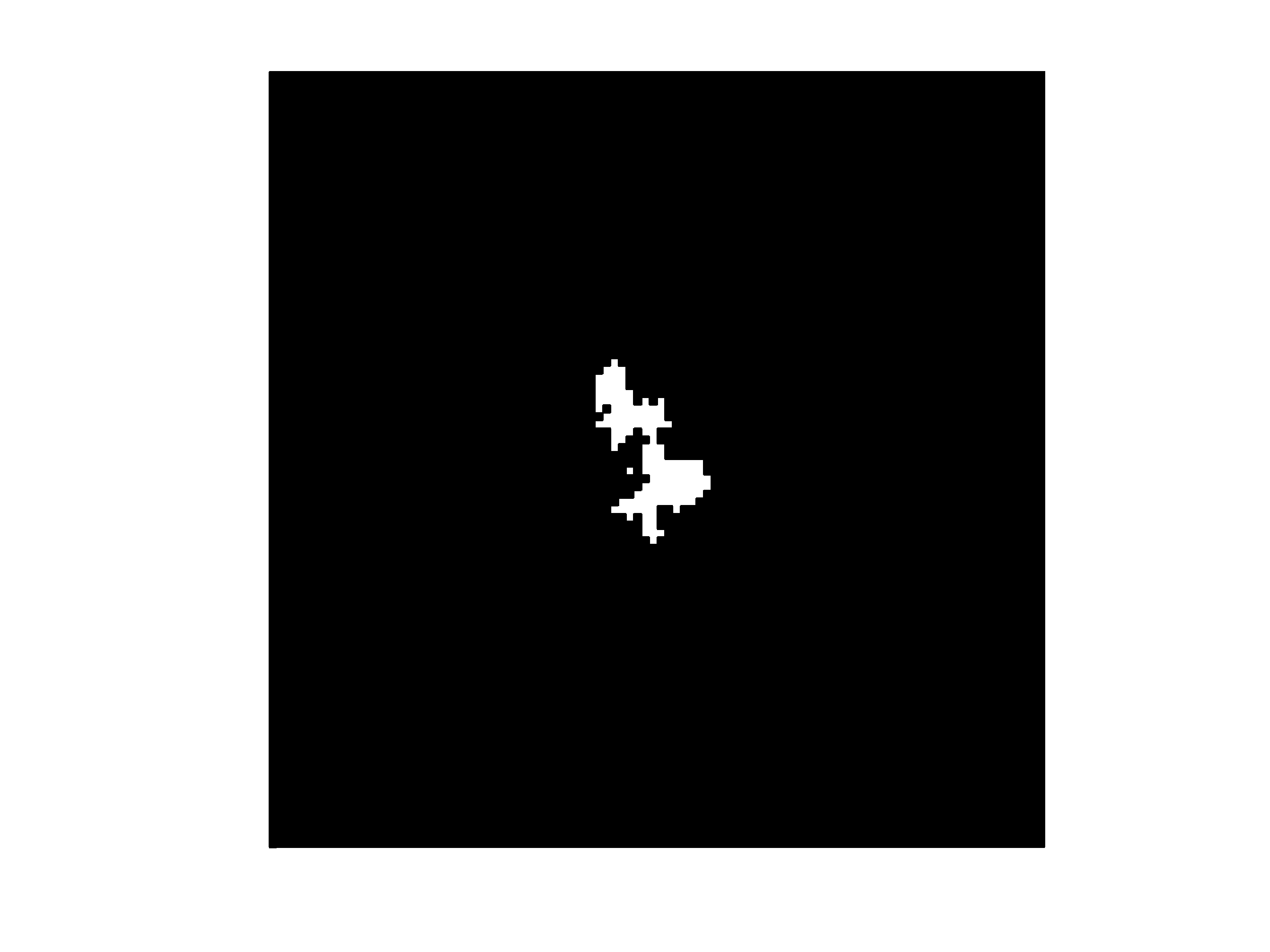}}

\subfigure[t=1000]{
\includegraphics[width=1.4in]{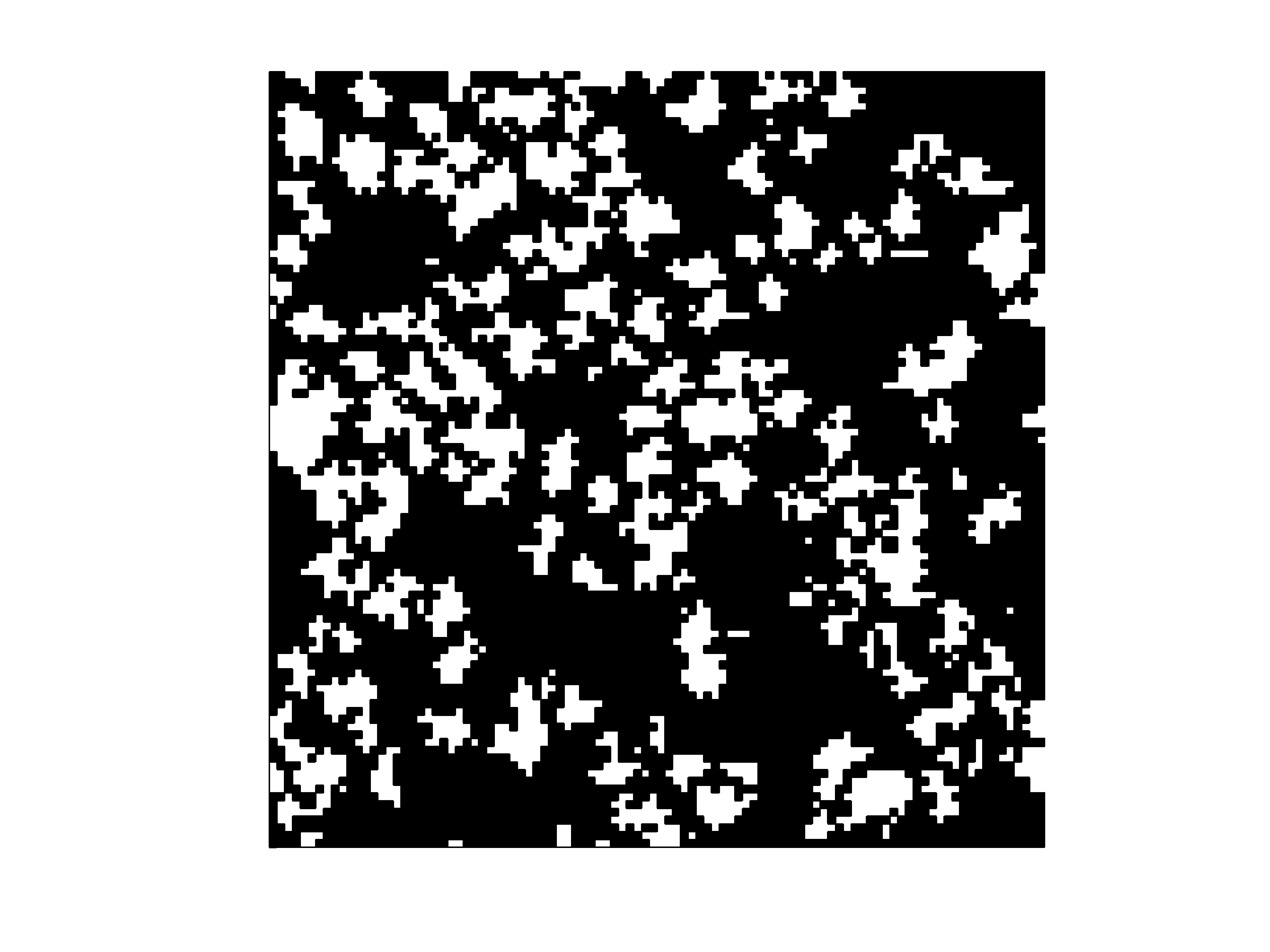}}
\centering
\hfill
\subfigure[$\rho(t)$]{
  \includegraphics[width=1.4in]{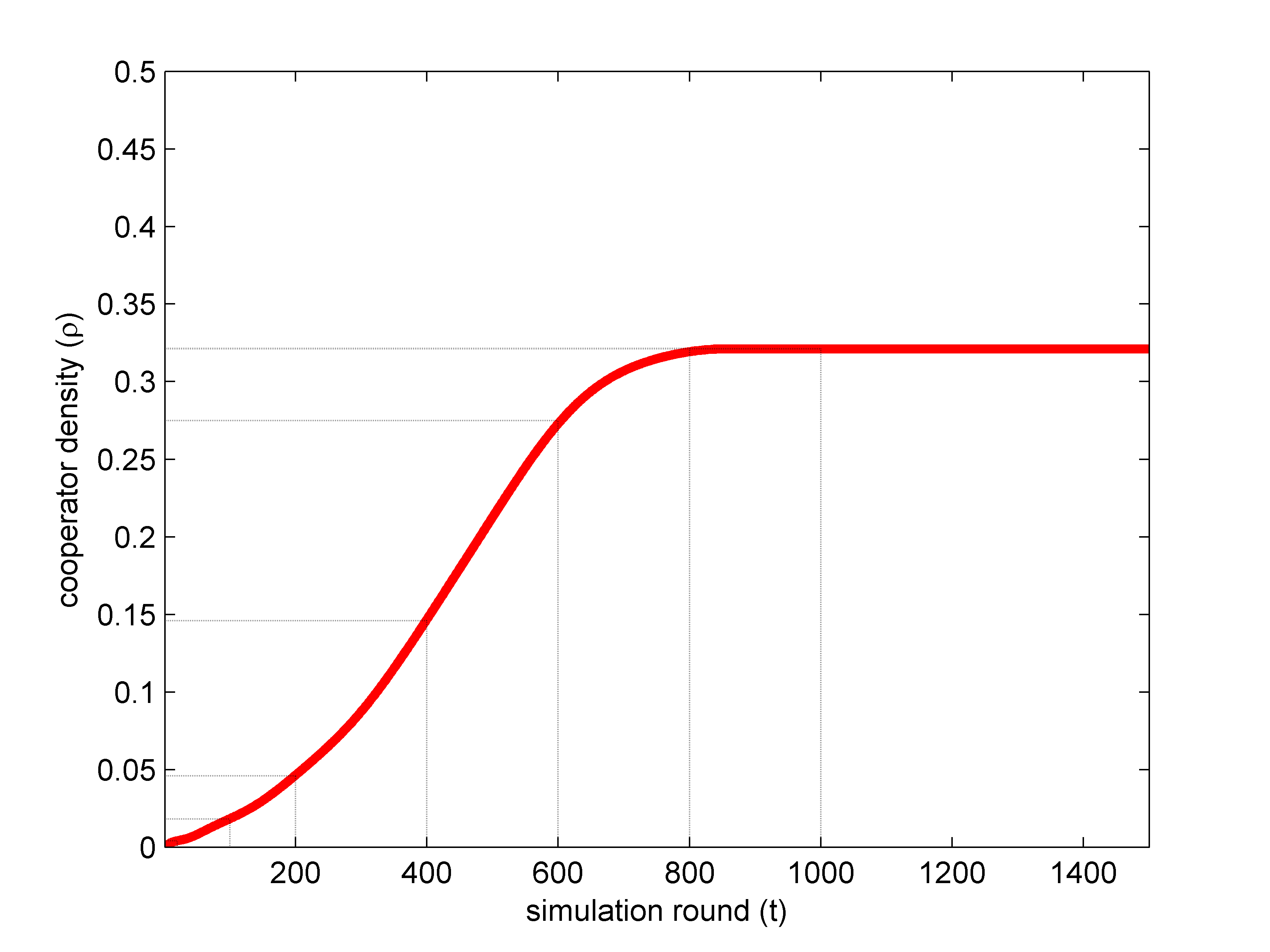}}
  \centering
\hfill
\subfigure[t=200]{
  \includegraphics[width=1.4in]{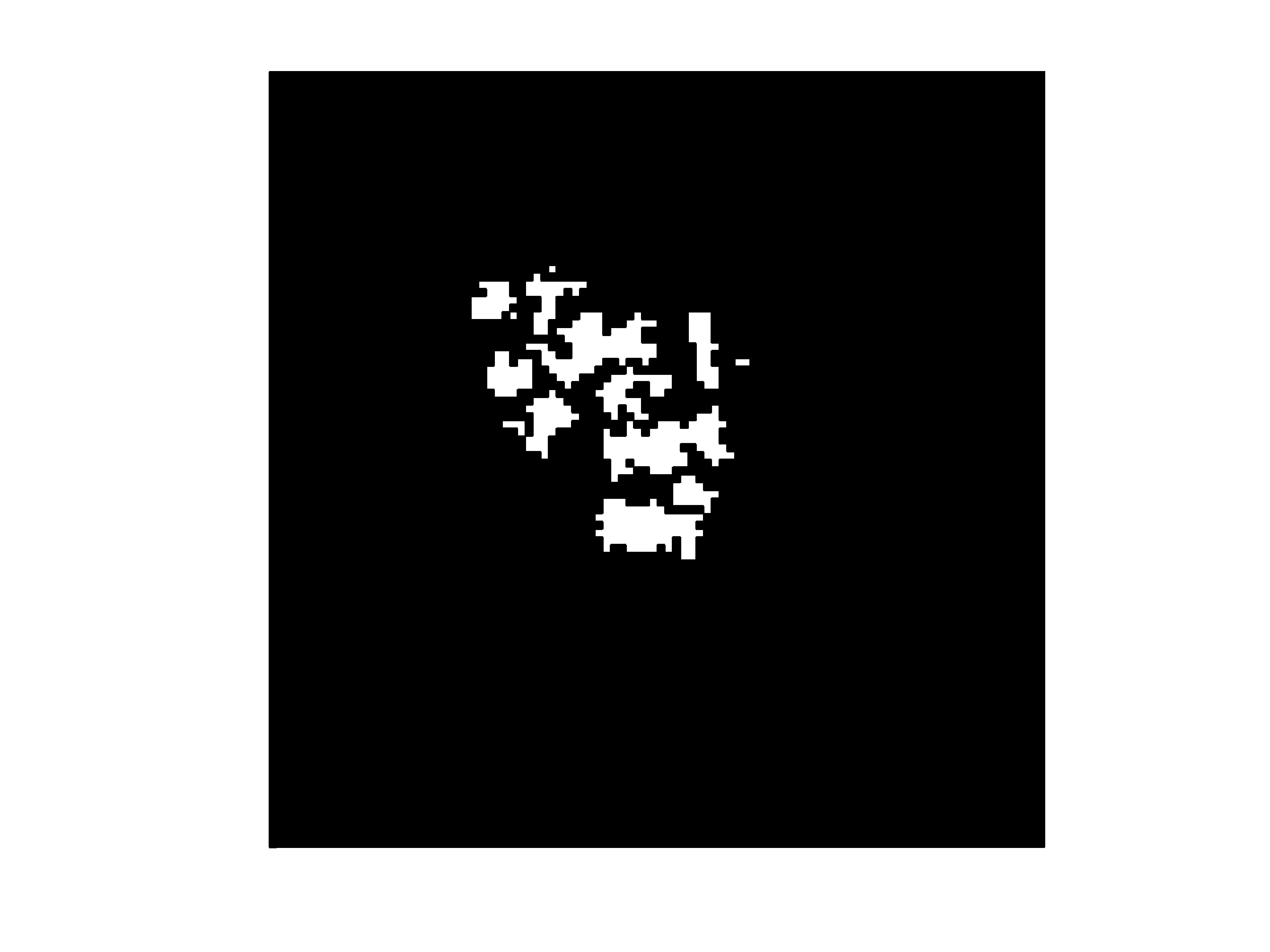}}

  \subfigure[t=800]{
\includegraphics[width=1.4in]{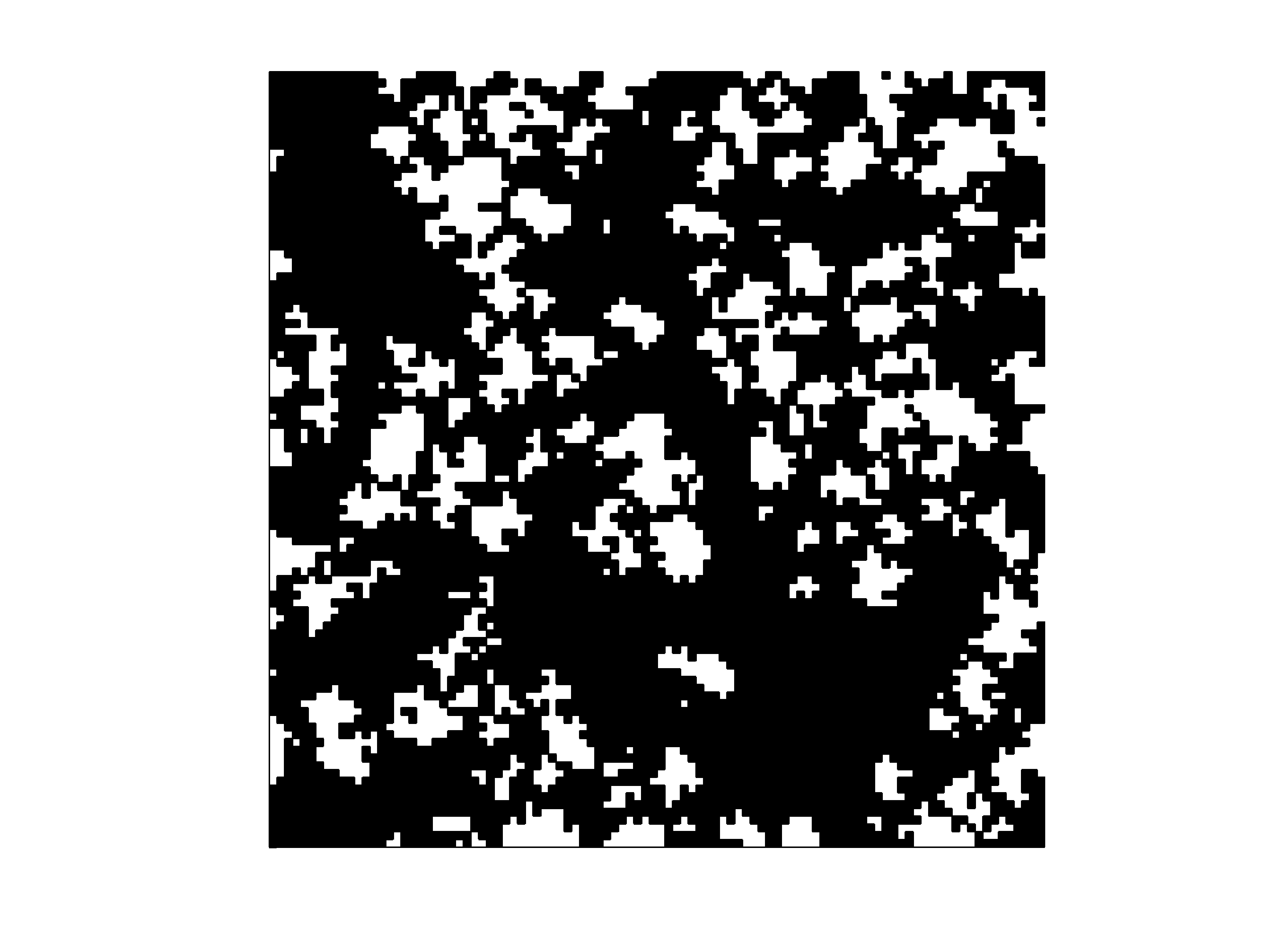}}
\centering
\hfill
\subfigure[t=600]{
  \includegraphics[width=1.4in]{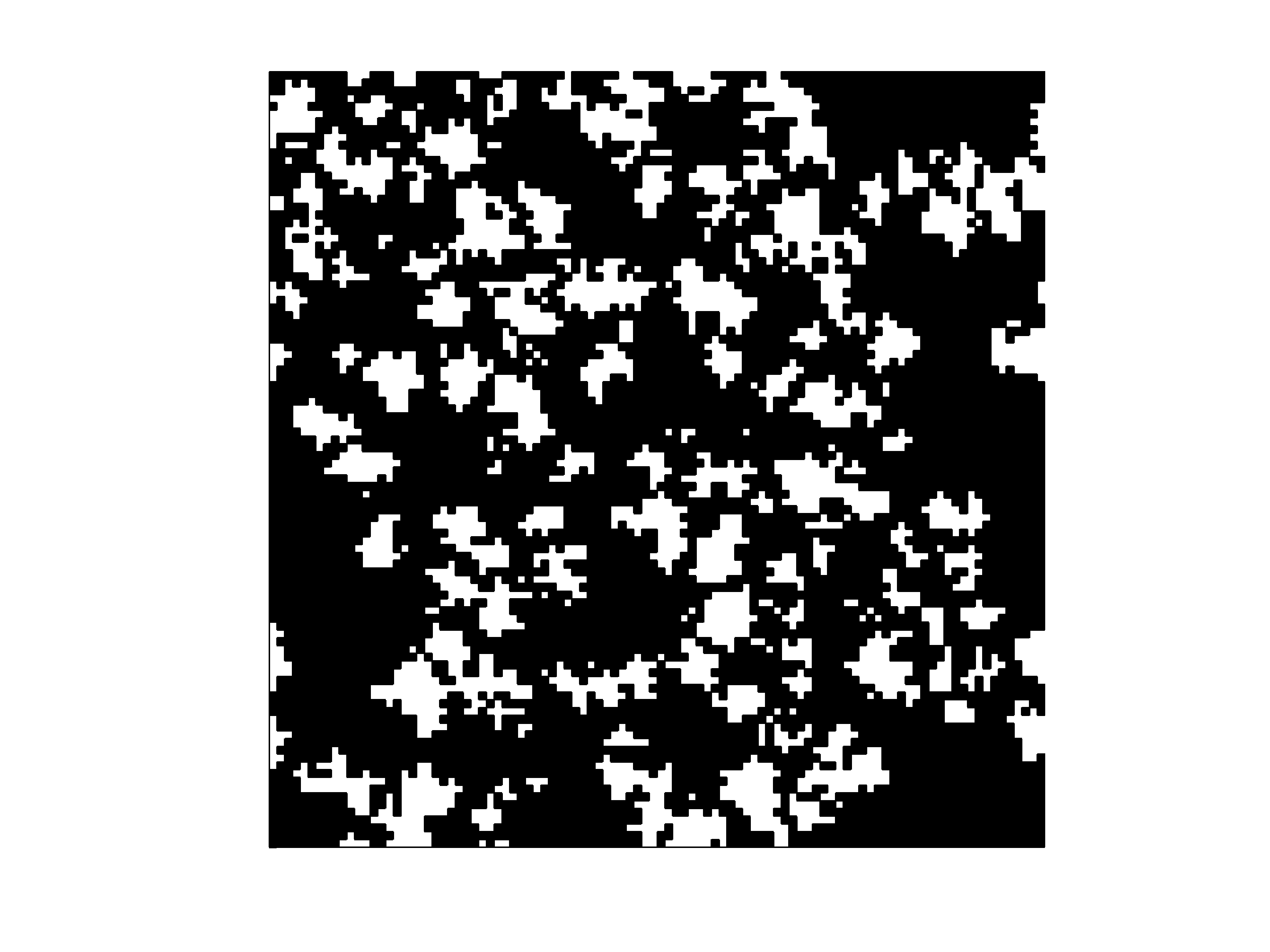}}
  \centering
\hfill
\subfigure[t=400]{
  \includegraphics[width=1.4in]{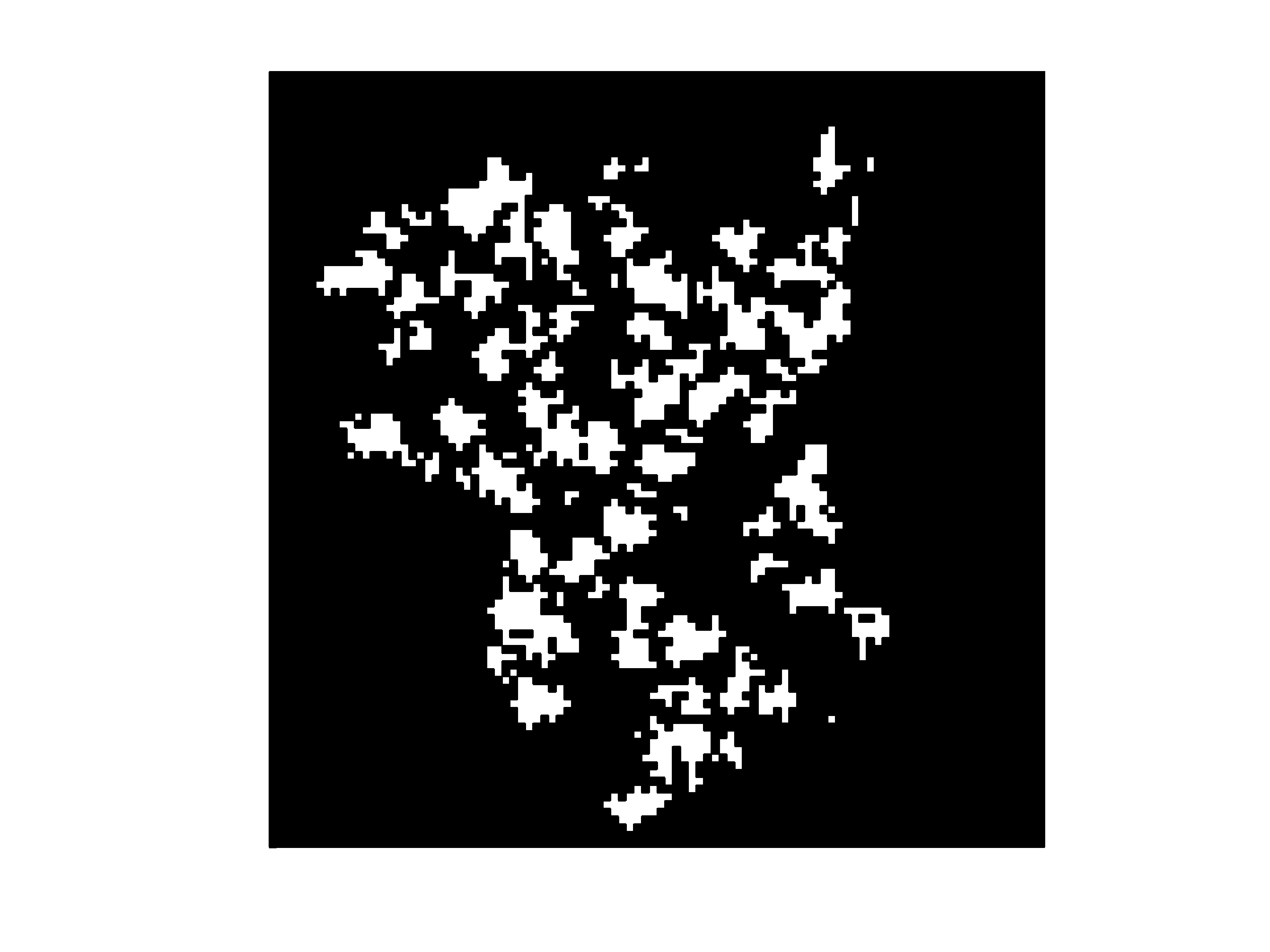}}

\caption{Snapshots of spatial distribution of cooperators (white boxes) and defectors (black boxes) at t=0, 20, 100, 200, 400, 600, 800 and 1000 when there is only one very small cooperative group in the center of the grid at the initial game moment. The middlemost subgraph shows the track of function $\rho(t)$, and the value of $b$ is still set to be 1.10.}
\end{figure*}
\par
Meanwhile, monte carlo strategy continues the inclusiveness of fermi rule and moran process for irrational behavior [27]. It considers the individual's game returns as his fitness for the society, and the higher the game returns, the more the one can adapt to the society.
In general, monte carlo rule is an organic combination of the current dynamic rules of individual policy adjustment, so that it not only makes full use of information, but also reflects the individual's bounded rational behavior and the ambivalence between the pursuit of high interest and high risk.
\par
Further, we investigate the role of update rules in the prisoner's dilemma game on the grid. In this paper, we mainly compare three classic update rules with monte carlo rule: unconditional imitation rule, replicator dynamics rule and fermi rule. Here, we briefly outline them:
\par
1. Unconditional imitation rule: during the evolutionary process, the individual would compare his own returns with those of all the neighbors and choose the game strategy with the highest returns as his strategy in next round of game.
\par
2. Replicator dynamics rule: during the evolutionary process, the individual $i$ randomly choose a neighbor $j$ for returns comparison. If $j$'s game returns $U_j$ is greater than his own game returns $U_i$, the individual $i$ would imitate $j$'s strategy with probability $p$ in the next game round:
\begin{equation}
p=\dfrac{U_j-U_i}{D\cdot \mathrm{max}(k_i,k_j)}
\end{equation}
Where $U_i$, $U_j$ represent the returns of individual $i$, $j$ in the previous game round respectively, and $k_i$, $k_j$ are the number of neighbors they have. Parameter $D$ is the difference between the largest and the smallest parameters in the game matrix (i.e. $b$).
\par
3. Fermi rule: during the evolutionary process, the individual $i$ randomly choose a neighbor $j$ for returns comparison, imitating $j$'s strategy with a certain probability which depends on the difference between the two:
\begin{equation}
p=\dfrac{1}{1+e^{(U_i-U_j)/\lambda}}
\end{equation}
where $\lambda$ represents the noise effect according to the rationality of individuals.
\begin{figure*}[!h]
\centering
\includegraphics[width=0.75\textwidth]{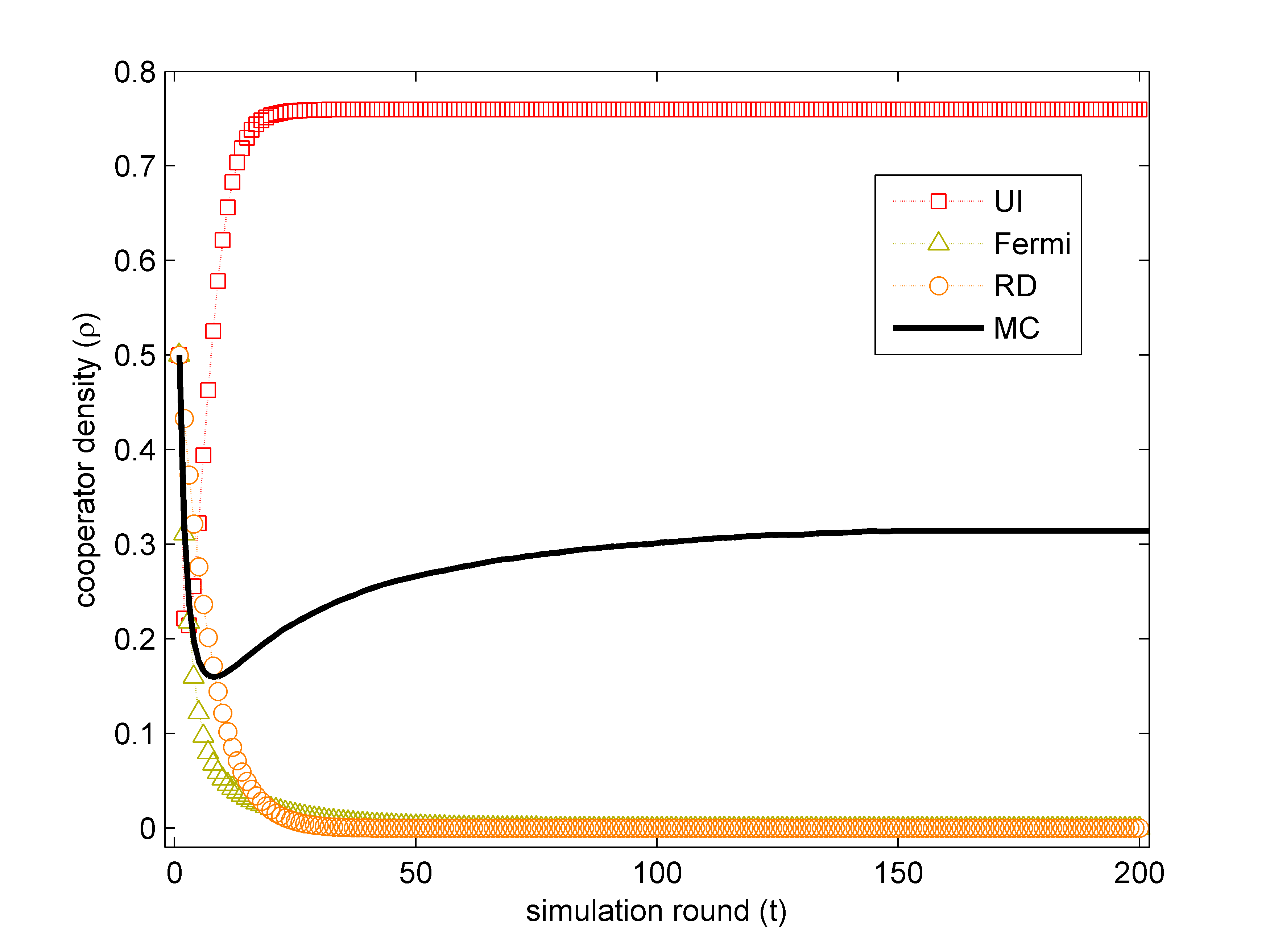}
\caption{Numerical simulation of cooperator density under monte carlo rule (solid line), replicator dynamics rule (circles), fermi rule (triangles) and unconditional imitation rule (squares) on the grid at b=1.10. 1000 simulations are averaged in each case.}
\end{figure*}
\par
We takes $\lambda$=0.0625 in fermi rule for the reason that, in this case the track of $\rho(t)$ is closest to those under monte carlo rule with the guidance of Mean Field Theory. This setting makes it more fair to compare the effects on cooperation level. Since individuals cannot judge the priority of the two strategies at the beginning, their initial strategies are all based on coin toss. That is, cooperators and defectors occupy 50\% of the grid, respectively. The result are showed in Fig. 3.
During the PD game, the $\rho$ tends to be stable promptly. We observe that the aggregate cooperation level between individuals is largely elevated under unconditional imitation rule or monte carlo rule, when compared to fermi rule and replicator dynamics. It is clearly that update rules play an important role in the evolutionary theory [19].
\begin{figure*}[!h]
\centering
\subfigure[(a) R=4\% ]{
  \includegraphics[width=1.4in]{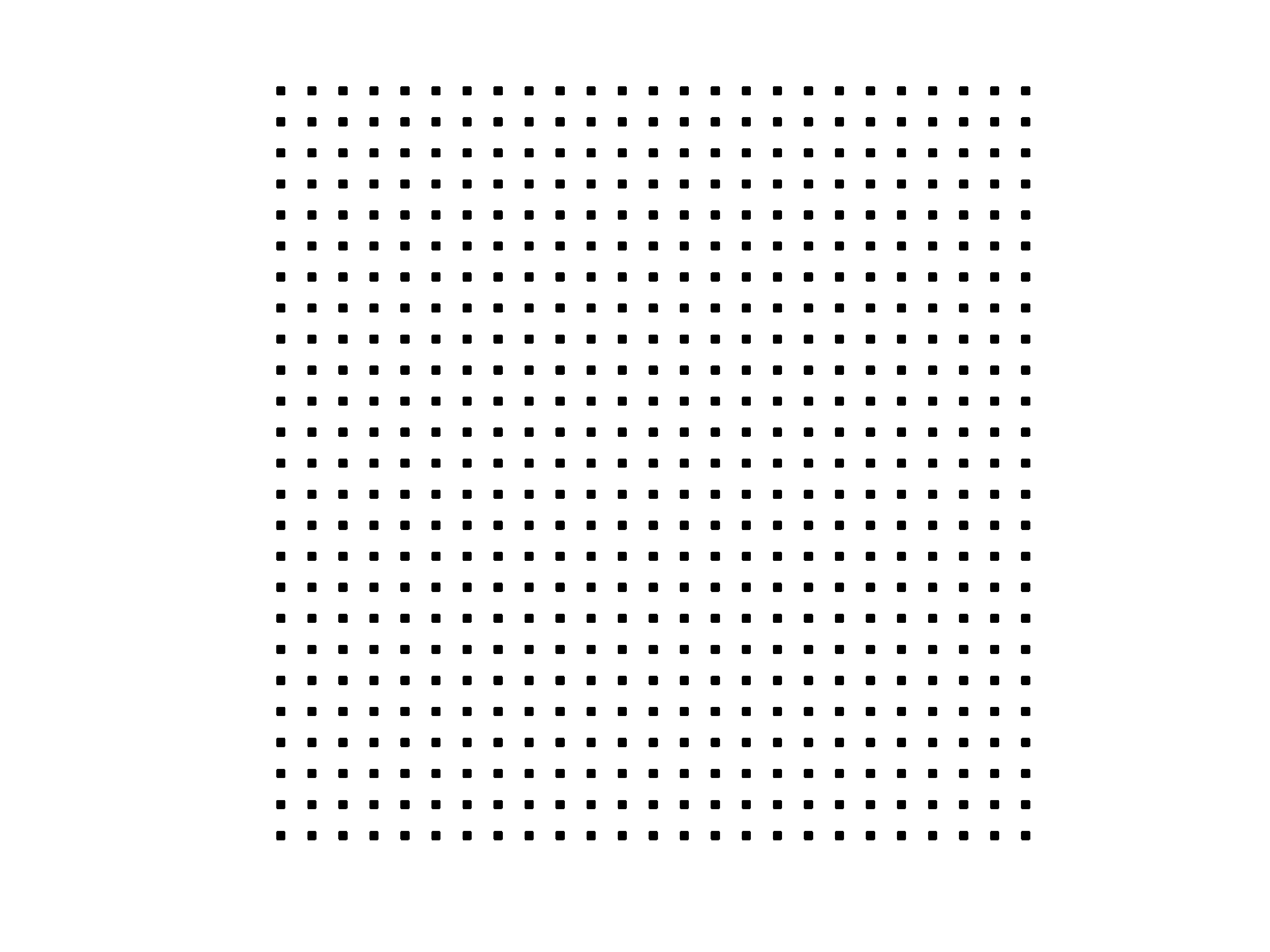}}
\centering
\hfill
\subfigure[UI]{
  \includegraphics[width=1.4in]{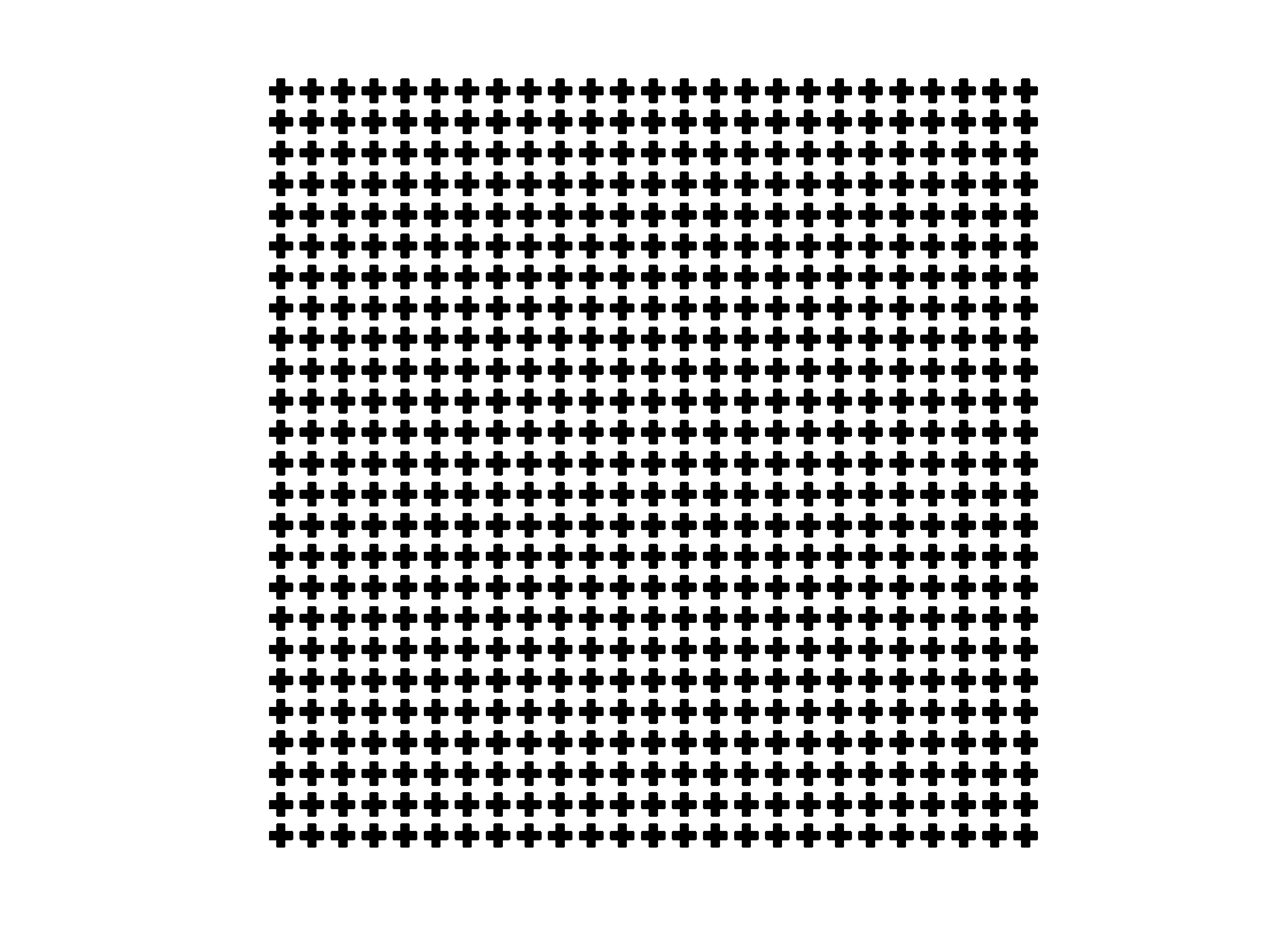}}
\centering
\hfill
\subfigure[MC]{
  \includegraphics[width=1.4in]{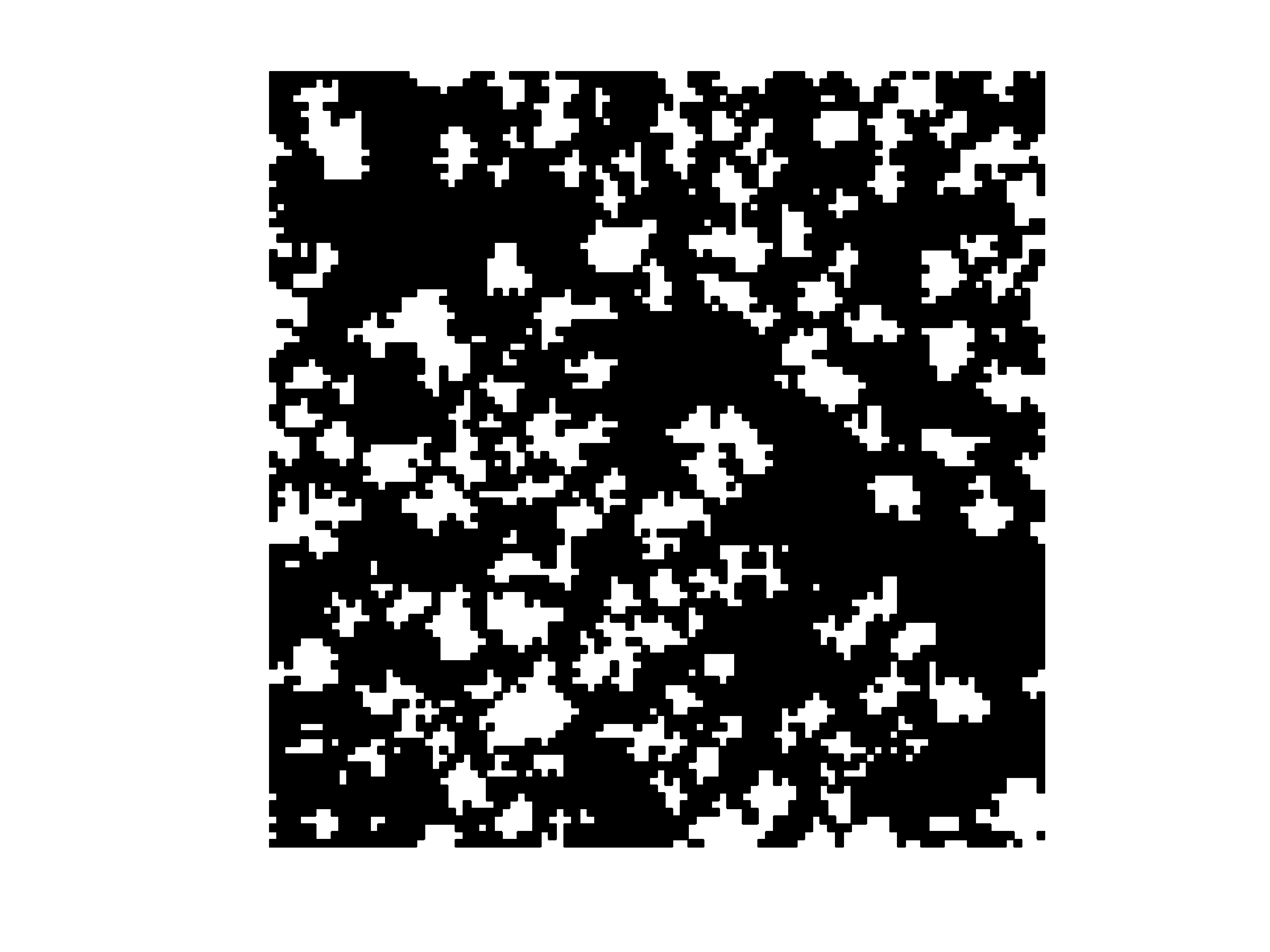}}
\subfigure[(b) R=6\%]{
\includegraphics[width=1.4in]{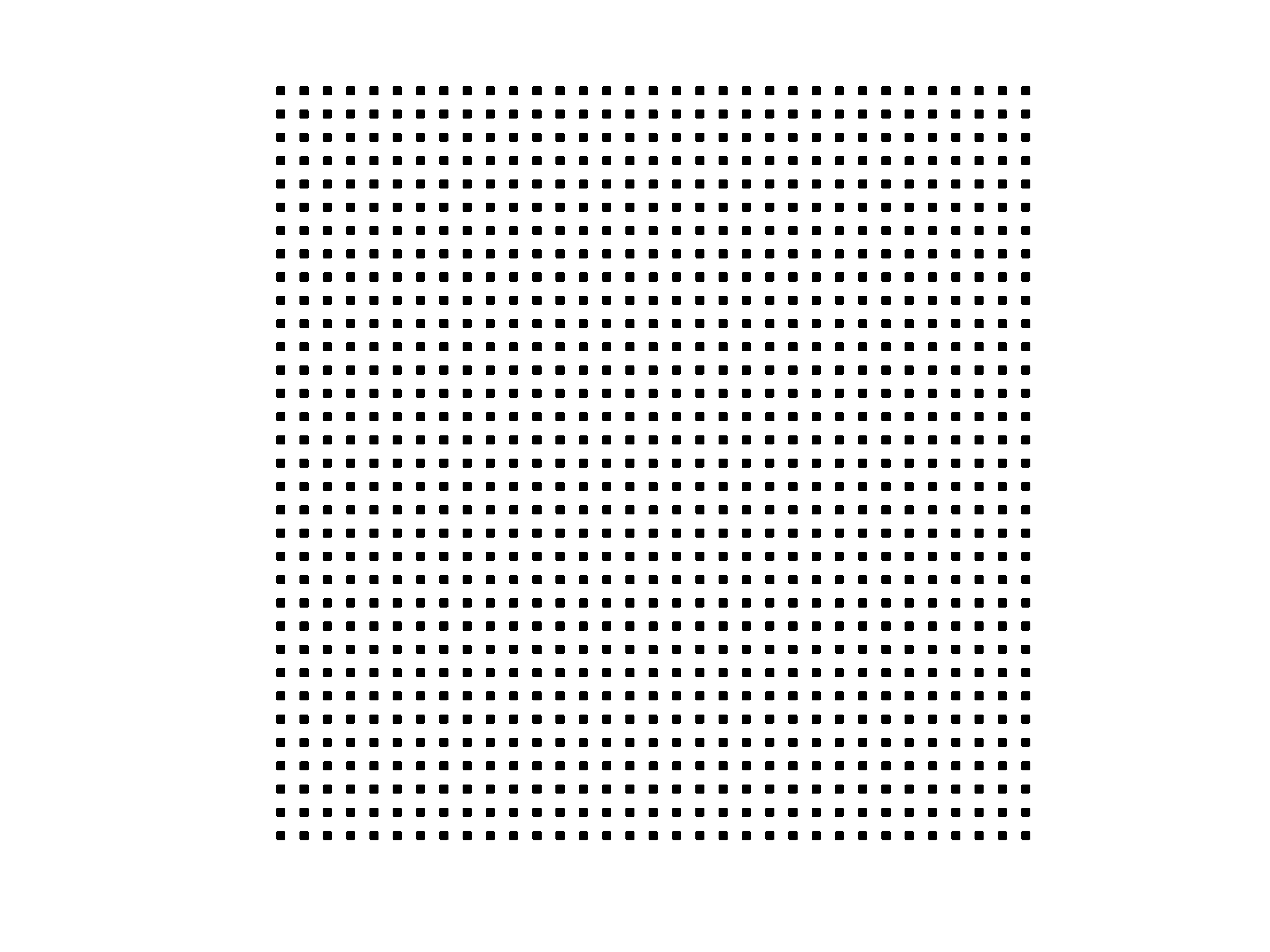}}
\centering
\hfill
\subfigure[UI]{
  \includegraphics[width=1.4in]{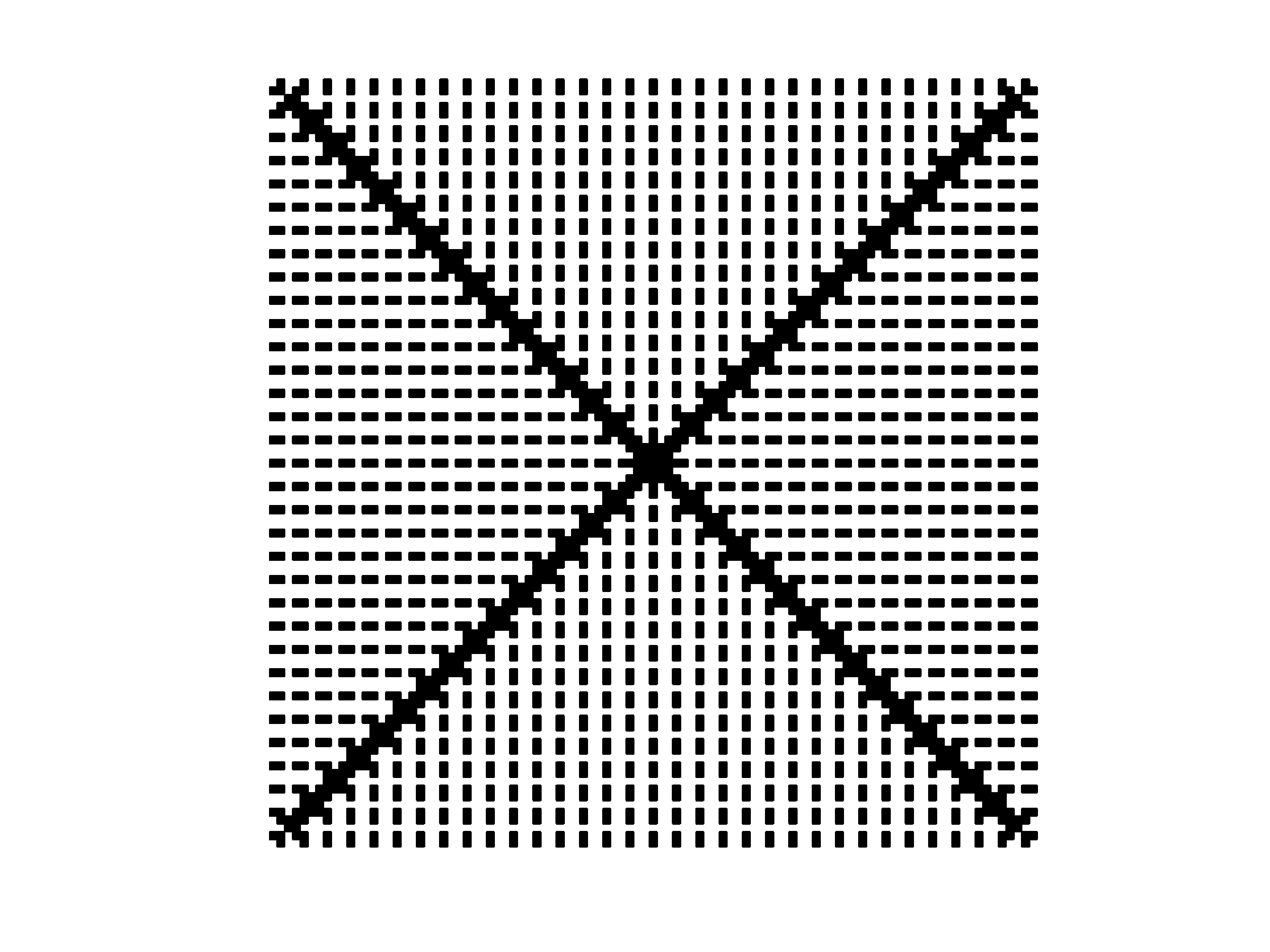}}
  \centering
\hfill
\subfigure[MC]{
  \includegraphics[width=1.4in]{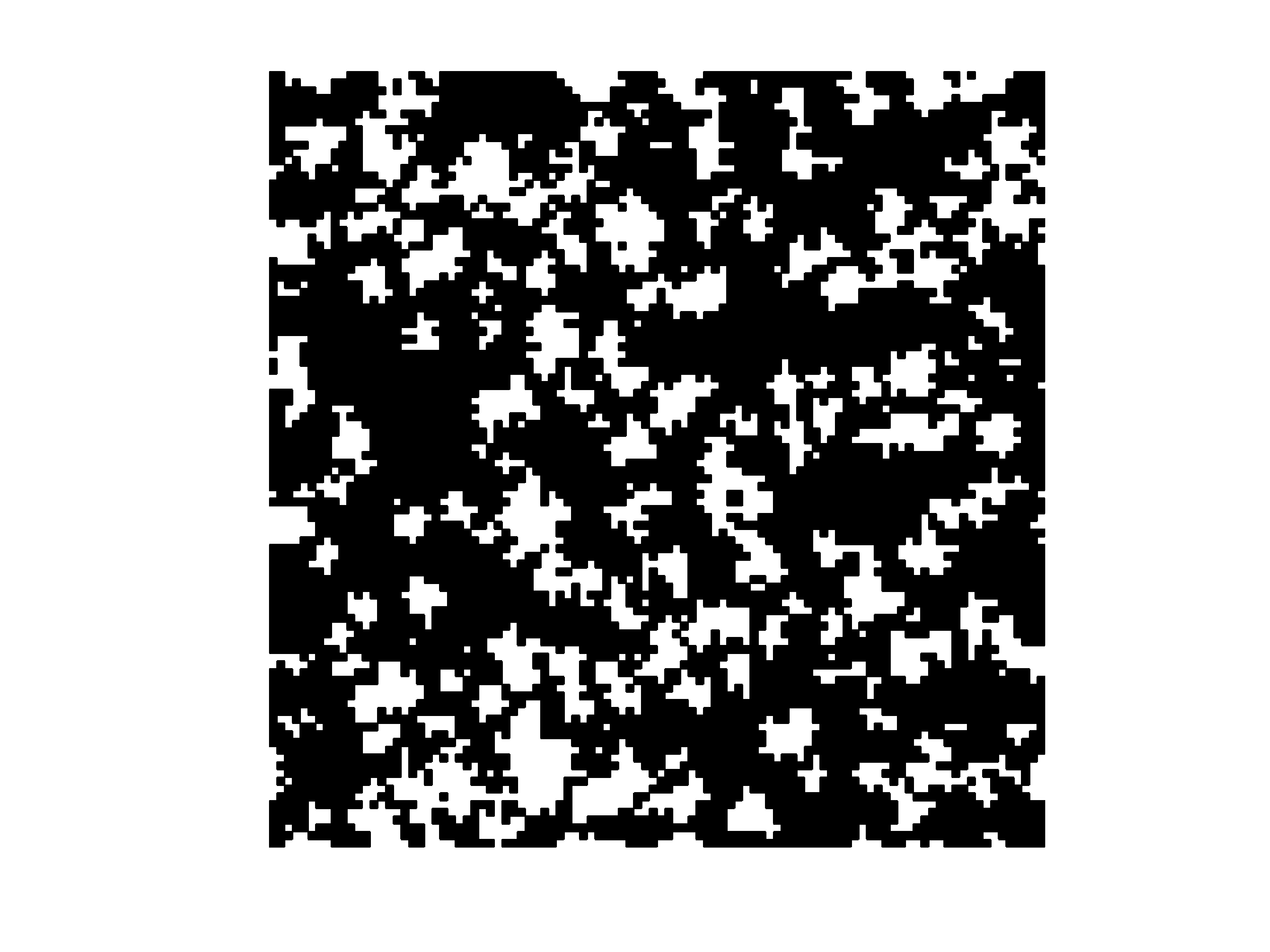}}

  \subfigure[(c) R=11\%]{
\includegraphics[width=1.4in]{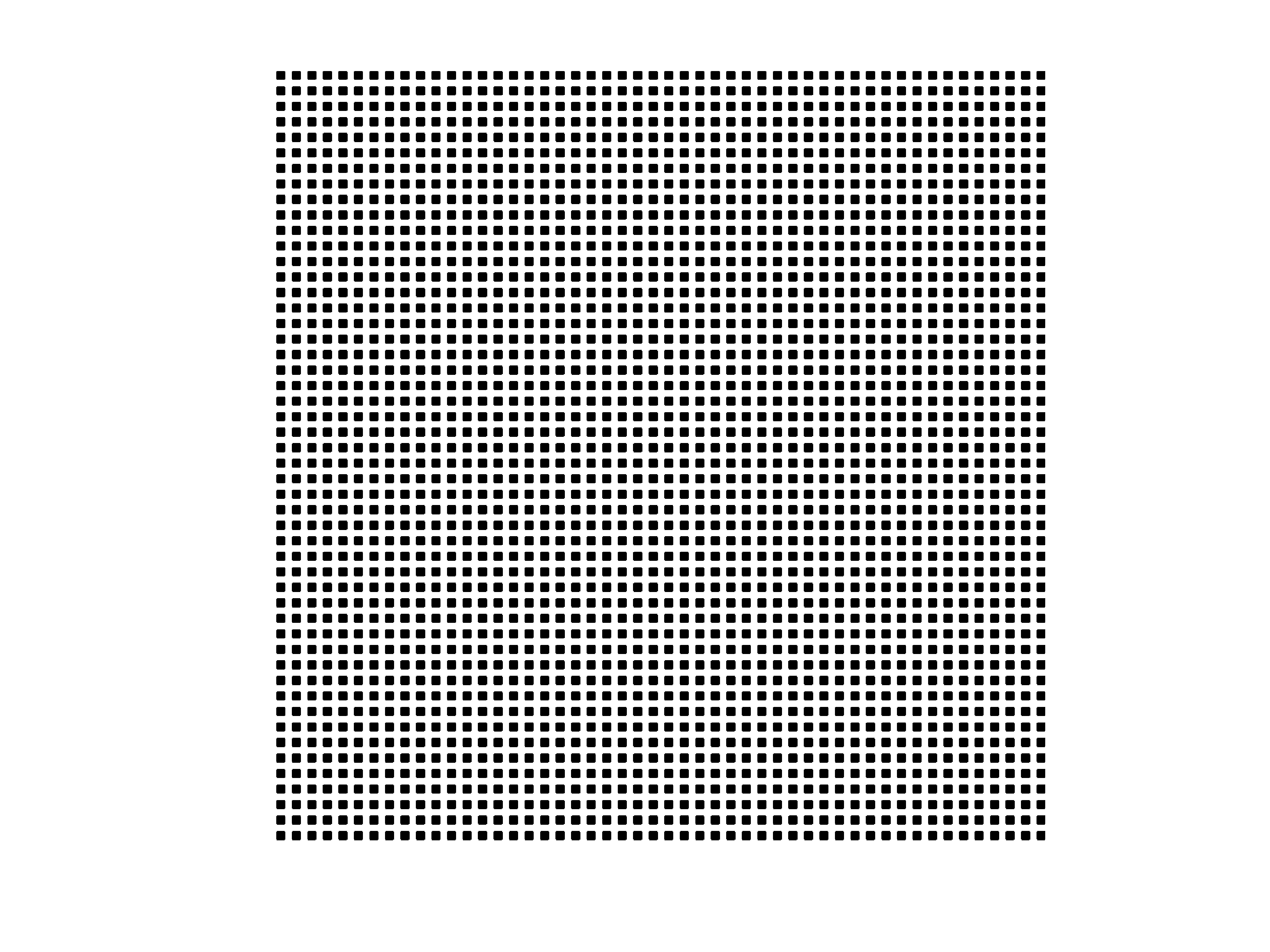}}
\centering
\hfill
\subfigure[UI]{
  \includegraphics[width=1.4in]{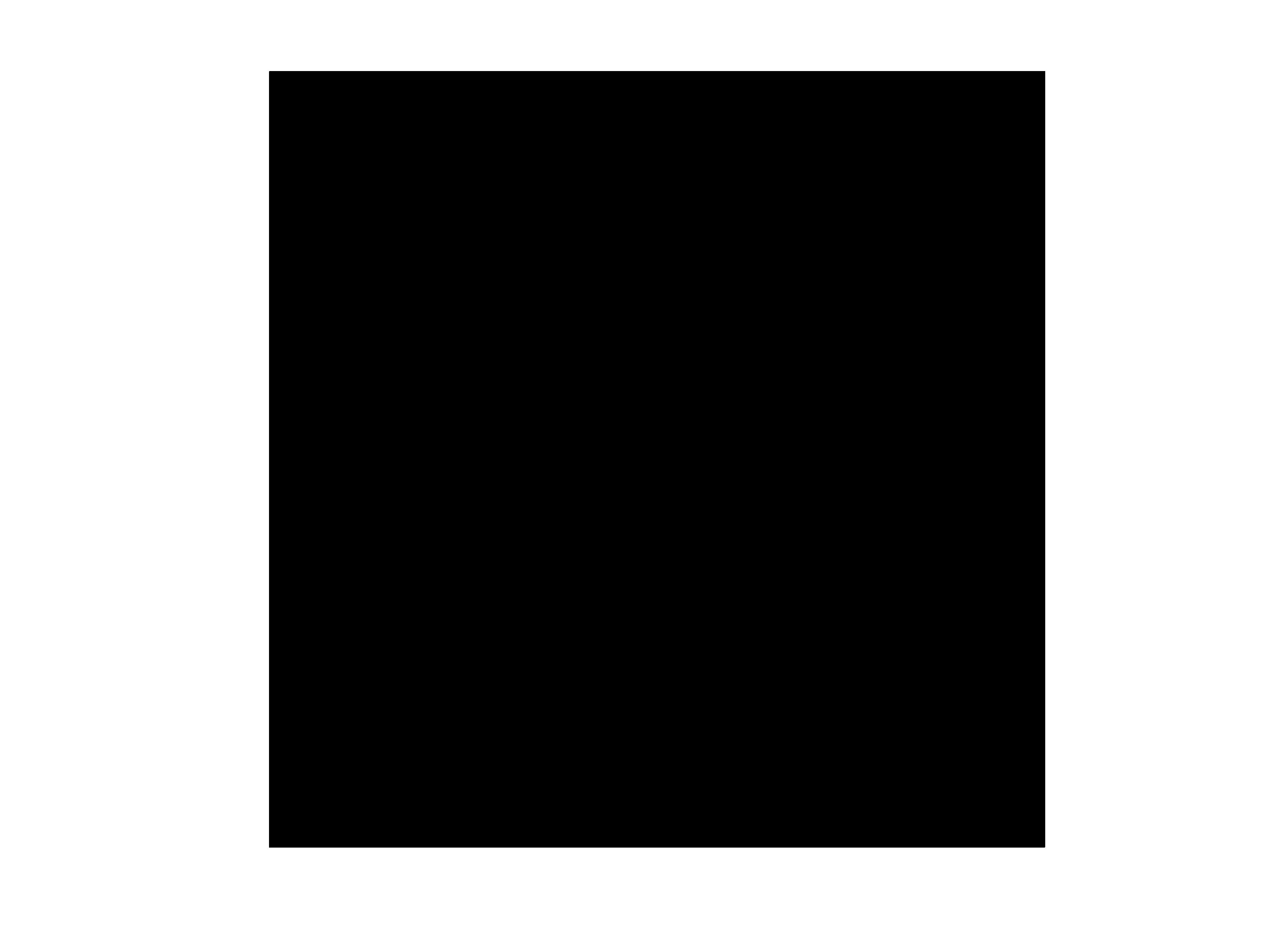}}
  \centering
\hfill
\subfigure[MC]{
  \includegraphics[width=1.4in]{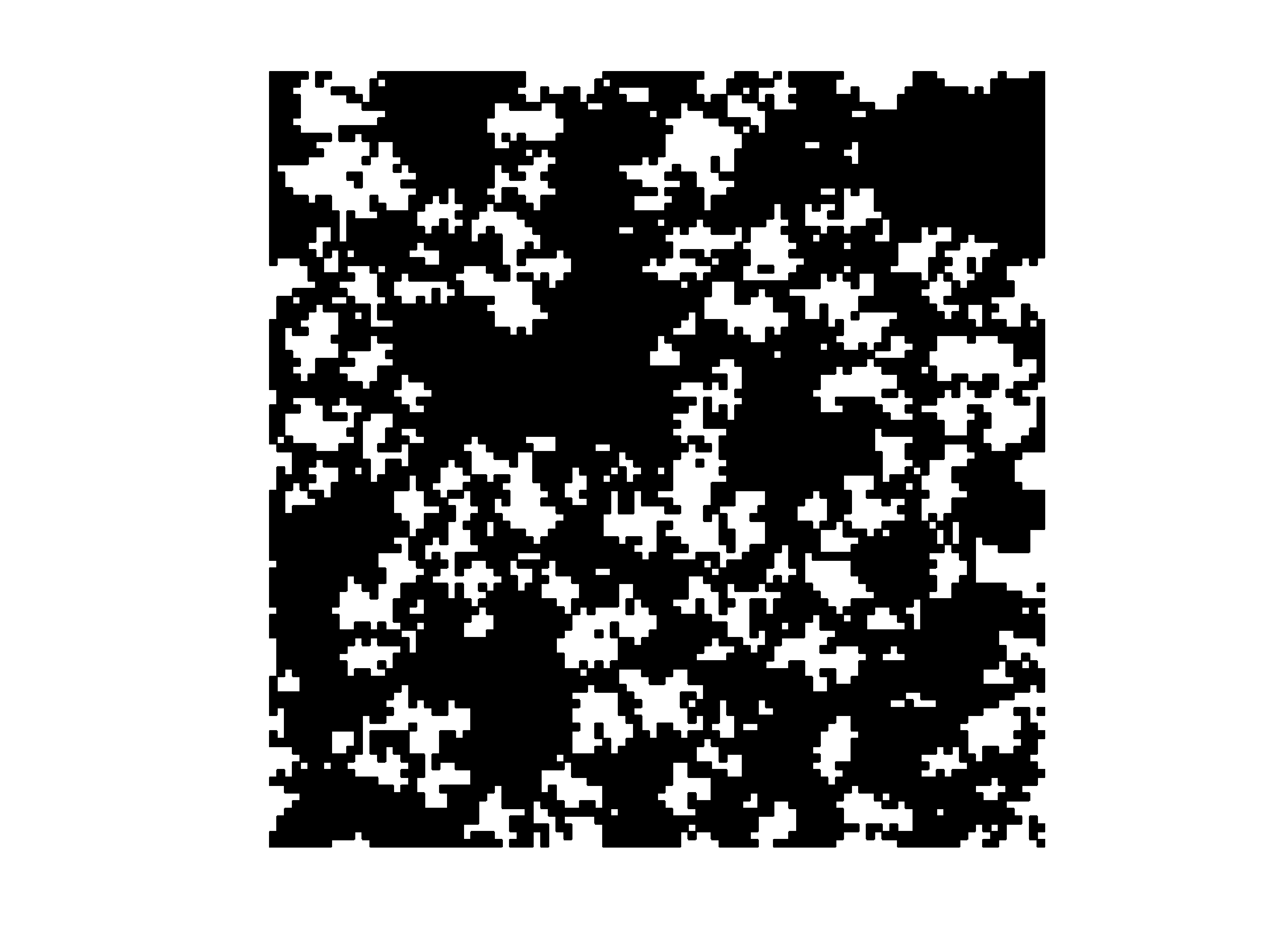}}

\caption{The influence of defective invasions on society. White boxes represent cooperators and  black boxes represent defectors. Each row corresponds to the different number of defectors ($R$), which are around 4\%, 6\%, 11\% of the society, respectively.
The first column shows the strategic distribution when the invasion occurs, the second column shows the game equilibrium results under the unconditional imitation rule, and the third column shows the game equilibrium results under the monte carlo rule. $N$ is set to be 10000 and $b$ is set to be 1.10. }
\end{figure*}
\par
Although cooperation level under unconditional imitation rule is higher than that of monte carlo rule, its mechanism of just imitating the best without any hesitation makes individuals no longer consider any risks behind the high-yields, so there is a reason to believe that it is not of robustness.
To confirm this fact, we consider an extreme situation. All the individuals on the grid are cooperators, and at some point, a group deliberately change its game strategy (defect). After that, we simulate enough time steps for accommodation of the defective invasion. The results are shown in Fig. 4. Each row corresponds to the different number of defectors ($R$), which are around 4\%, 6\%, 11\% of the society, respectively. The first column shows the strategic distribution when the invasion occurs, the second column shows the game equilibrium results under the unconditional imitation rule, and the third column shows the game equilibrium results under the monte carlo rule. As the invasion of the defection, defectors' neighbors did not hesitate to imitate the defective strategy for higher returns under unconditional imitation, formed the defective core, which could not update its own strategy, resulting in a significant reduction in cooperation level.
However, under monte carlo rule, individuals not only seek for high returns, but also concern the high risks behind the high returns and would make a balance between the two. So no matter how serious the invasion, the cooperation will be balanced over time ($\rho=0.32$), not extinct.
It is overwhelmingly clear that the robustness of monte carlo rule is significantly better than unconditional imitation rule.
\par
Next, we investigate the fraction of cooperators ($\rho$) as a function of the temptation parameter ($b$) under monte carlo rule on the grid since $\rho(b)$ is a pretty meaningful quantity in evolutionary games. The results are showed in Fig.5(a). The monte carlo rule always performs a cooperative society when  temptation is low. As $b$ increases, the balanced pattern has been broke. On the border, the internal support can not conquer the loss caused by defectors, and the cluster begins to collapse layer by layer until a new balance appears. The cooperators go extinct ultimately at $b$=1.31, for the reason that the huge temptation prevents any form of clusters from overcoming the loss on the border.
\begin{figure*}[h]
\centering
\subfigure[(a) $\rho(b)$]{
  \includegraphics[width=2.2in]{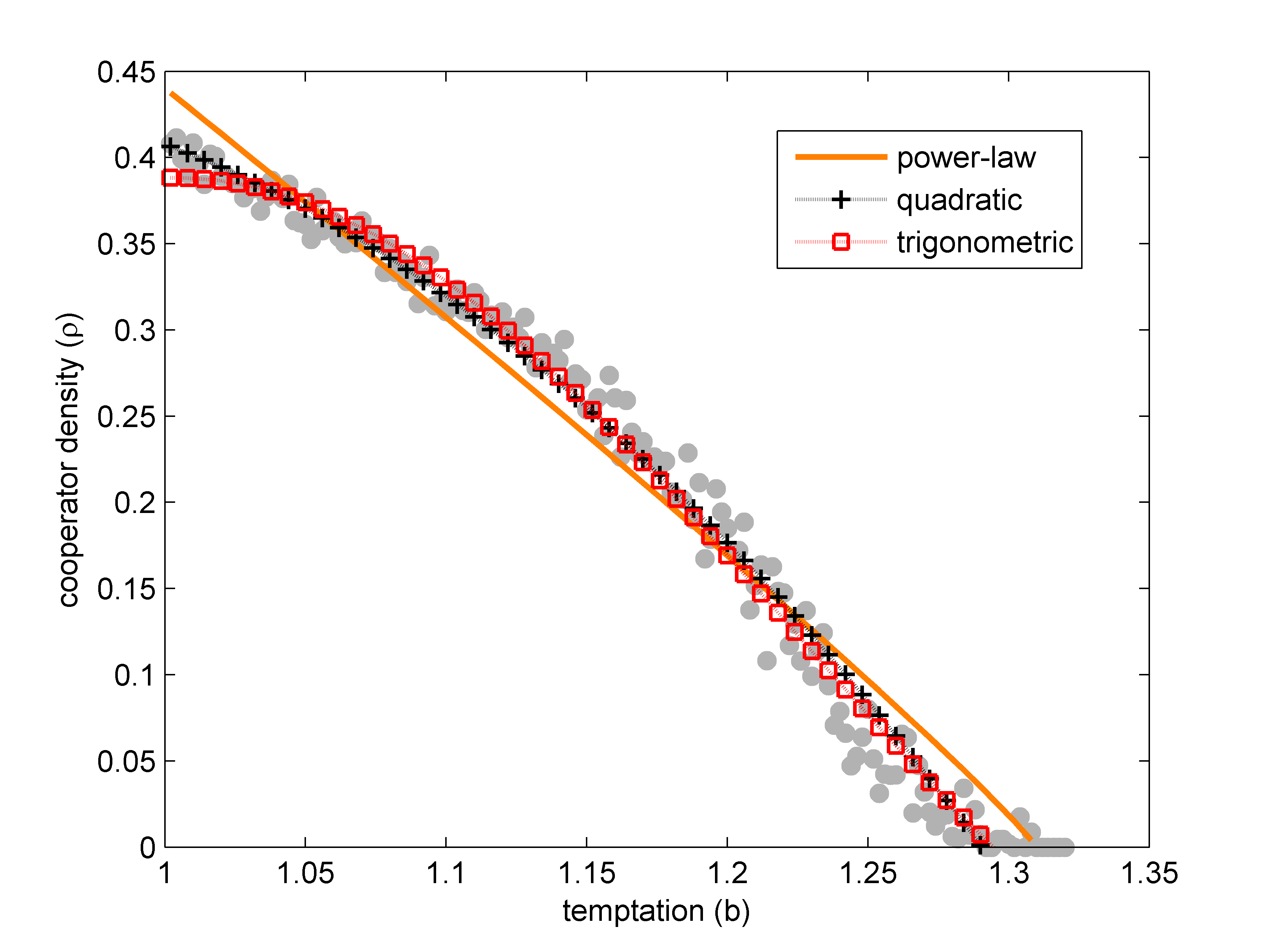}}
\hfill
\centering
\subfigure[(b) returns comparison]{
  \includegraphics[width=2.2in]{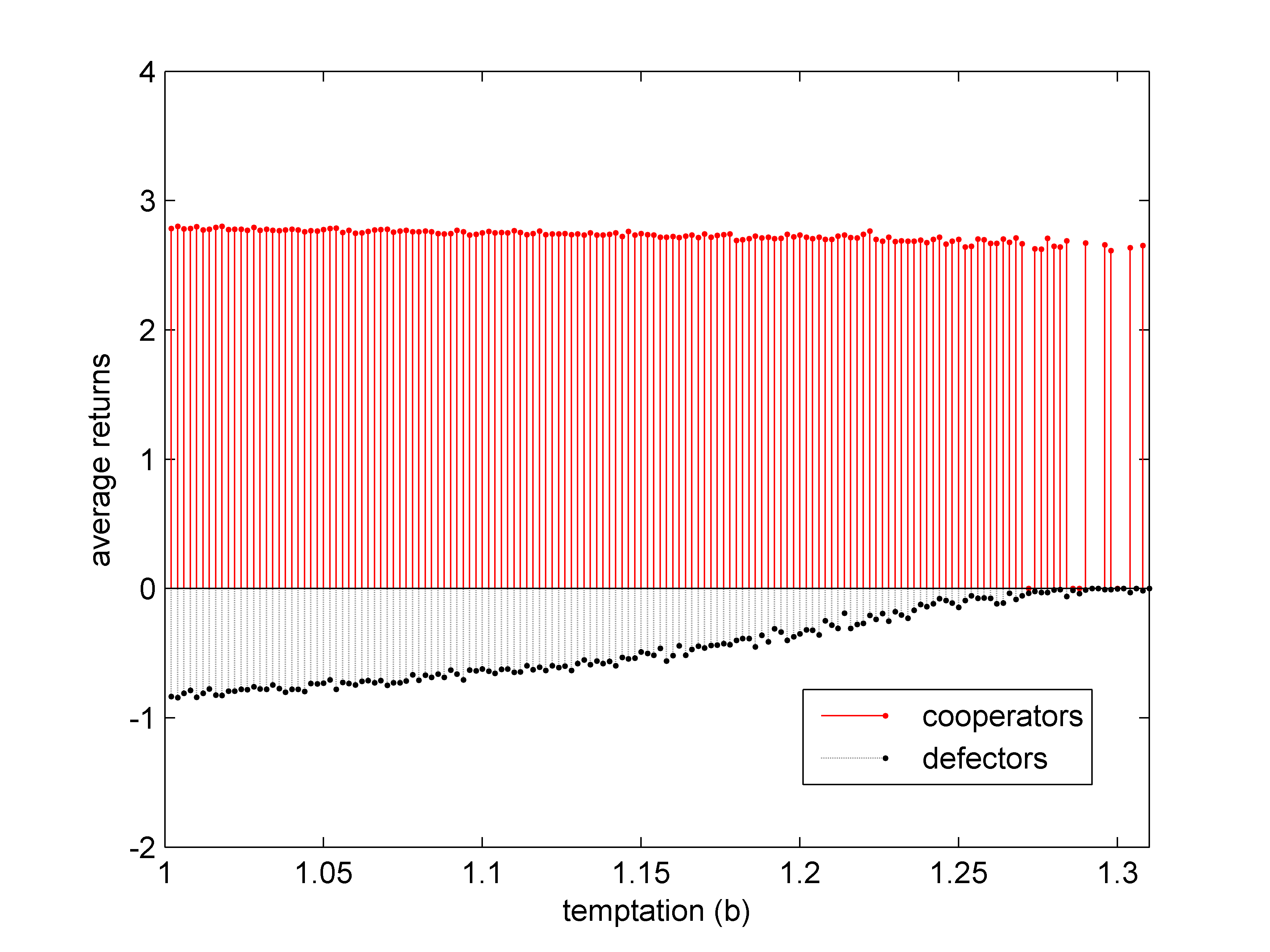}}
\caption{(a) The effect of power law fitting (solid line), quadratic fitting (pluses) and trigonometric fitting (squares) on the curve $\rho(b)$. For easy comparison, the original data expressed by gray dots are also given in the figure. (b) The relationship between average returns of individuals and the temptation parameter ($b$). 1000 simulations are averaged in each case.}
\end{figure*}
\par
According to Szabo's outstanding work, the cooperator density data refer to a power-law behavior in prisoner's dilemma game under the fermi rule on the grid, that is $\rho \propto (b_{cr}-b)^\beta$ [20], in which $b_{cr}$ is the threshold of the disappearance of cooperation, once $b>b_{cr}$, then all the cooperators would go extinct.
With the guidance of the goodness of fit ($R$) for nonlinear regression and Root Mean Squared Error ($RMSE$) displayed below:
\begin{equation}
R=1-(\sum{(y-\hat{y})^2} / \sum{y^2})^ {1/2}
\end{equation}
\begin{equation}
RMSE=((\sum{(y-\hat{y})^2})/n)^{1/2}
\end{equation}
where $y$ represents the original data, $\hat{y}$ is on behalf of the fitting data, and $n$ is the length of $y$.
We find that power-law fitting originally performing very well under the fermi rule is unsatisfactory under monte carlo rule, which can be seen in Fig.5(a). The solid line represents the best result of the power-law fitting, where $b_{cr}$=1.31 and $\beta$=0.923. After further observing the original data, we choose quadratic and trigonometric curves to fit them. Their expressions can be respectively indicated as:
\begin{equation}
y=ax^2+bx+c \ \ \mathrm{and} \ \ y=A\mathrm{sin}(\omega x + \varphi)+d
\end{equation}
\par
\noindent
The fitting effects are also showed in Fig.5(a), and the specific fitting value
can be seen in Table 2.
The pluses represent the result of quadratic fitting where the most optimal parameter are $a$=-2.7363, $b$=4.8644 and $c$=-1.7205. The squares represent trigonometric's fitting effect where the best parameters are $A$=-0.2568, $w$=7.2462, $\varphi$=-2.5603 and $d$=0.1314. The fitting effects of the two are better than that of power-law fitting based on the smaller $RMSE$ and the better $R$. Furthermore, trigonometric fitting has a better result than quadratic fitting, so we say that the cooperator density data refer to a trigonometric behavior under the monte carlo rule on the grid, that is $\rho \propto A\mathrm{sin}(\omega x + \varphi)+d$.
\begin{table}[!h]
\caption{The optimal solution of thr three fitting methods.}
\label{tab:2}
\begin{tabular}{llll}
\hline\noalign{\smallskip}
Fitting method & Root Mean Squared Error & Goodness of Fit \\
\noalign{\smallskip}\hline\noalign{\smallskip}
power-law & 0.2608 & 0.9052 \\
quadratic  & 0.0168 & 0.9356\\
trigonometric  & 0.0149 & 0.9428\\
\noalign{\smallskip}\hline
\end{tabular}
\end{table}
\par
We have also observed the relationship between average returns of individuals and the temptation parameter $b$. The results are showed in Fig.5(b). Macroscopically, the strategy of cooperation is obviously better than defection due to the better returns. It is worth noting that the average returns of cooperators is insensitive to the temptation parameter $b$. Furthermore, it is surprised to find that the returns of the defectors decreases as the growth of $b$, for the reason that excessive defectors gatherings have greatly reduced the success of defection. It is clear that cooperation makes individuals live better in society.
\par
\begin{figure*}[htb]
\centering
\subfigure[(a) Sensitivity of $\rho$ to $\rho_0$]{
  \includegraphics[width=2.2in]{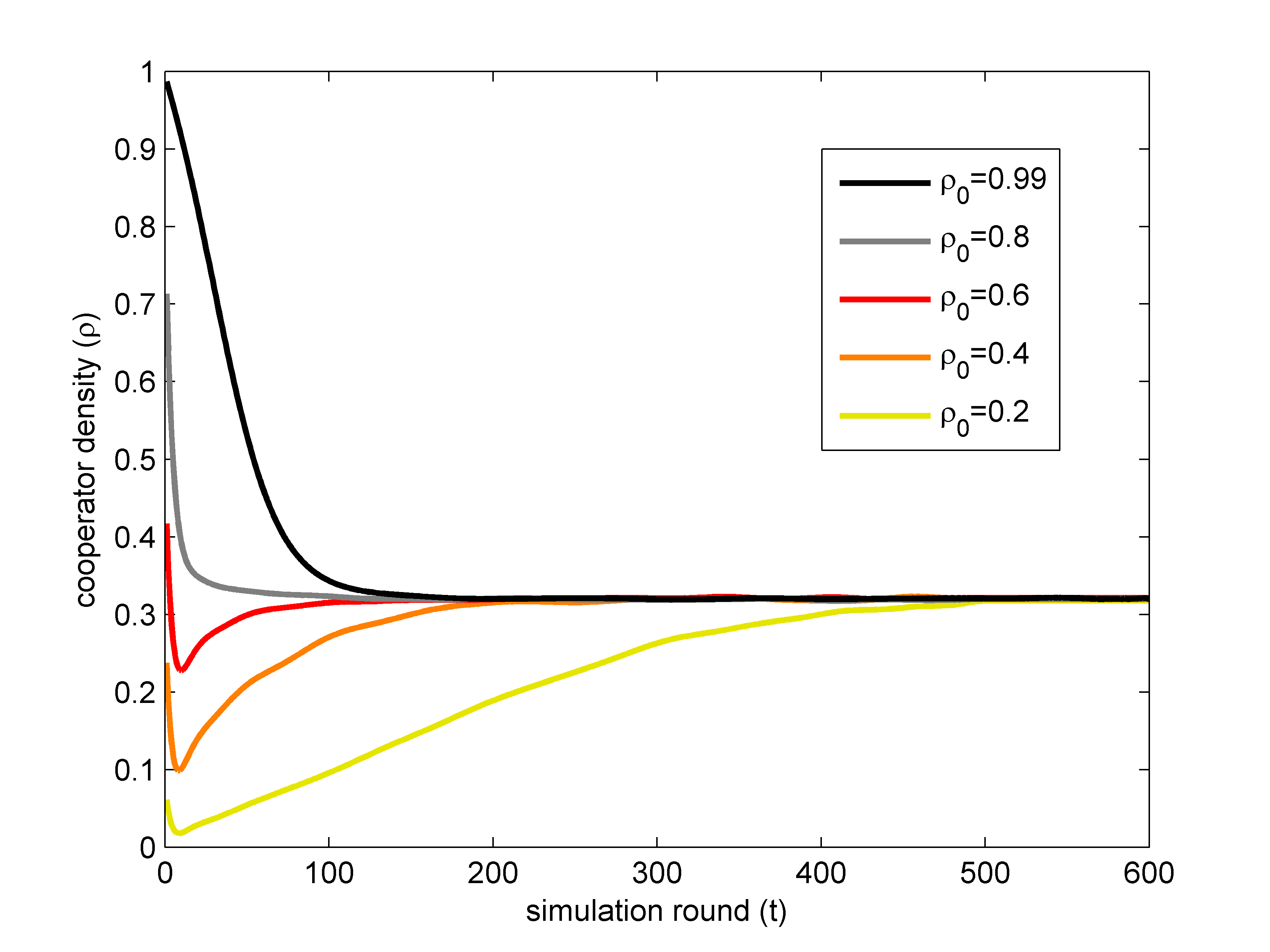}}
\hfill
\centering
\subfigure[(b) Sensitivity of $\rho$ to $N$]{
  \includegraphics[width=2.2in]{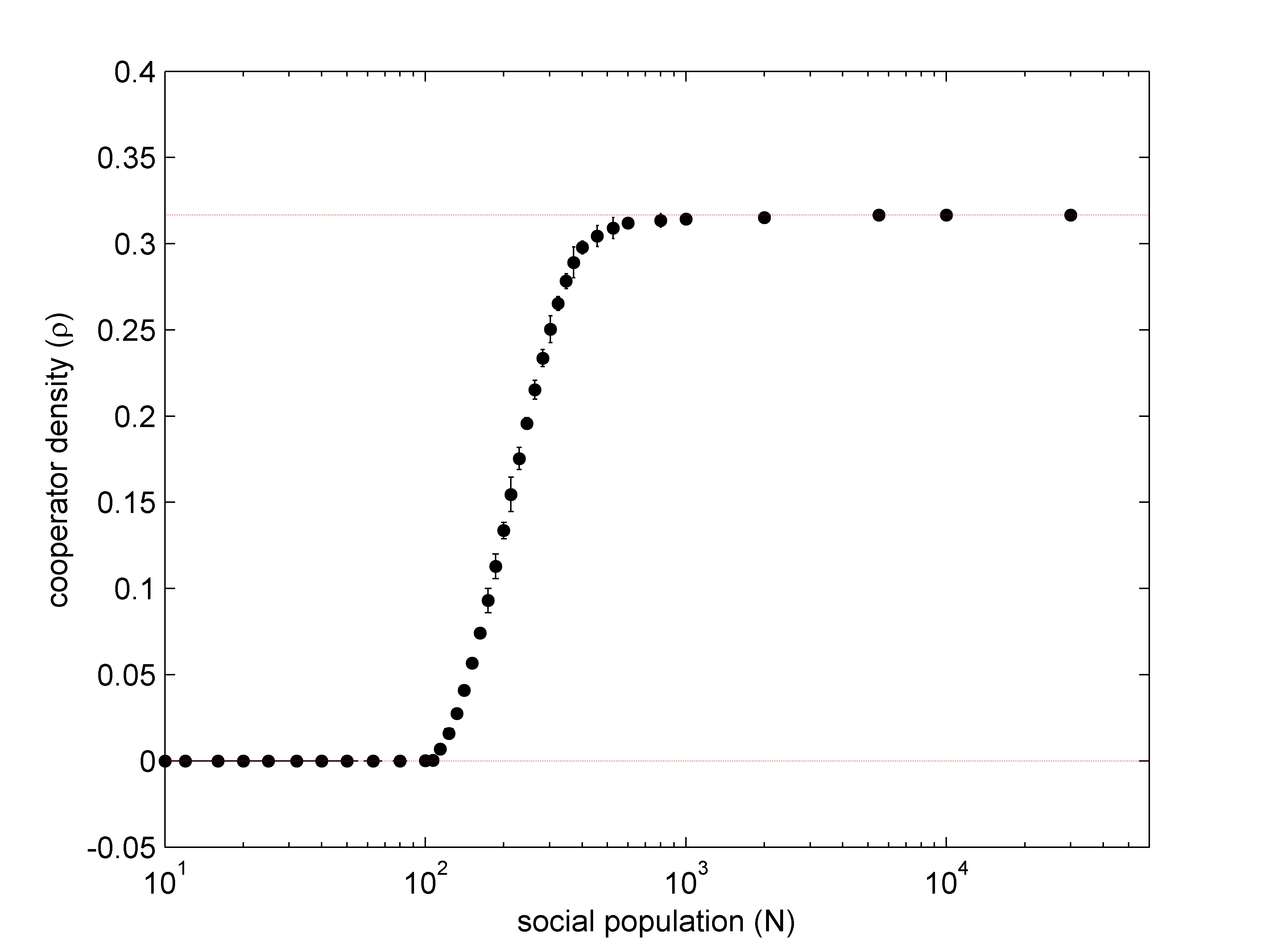}}
\caption{(a) The numerical simulation results for the density of cooperator ($\rho$) as the game progresses for different initial density of cooperator ($\rho_0$) on the grid at $b$=1.10. From bottom to top, the $\rho_0$ is set to be 0.2, 0.4, 0.6, 0.8 and 0.99, respectively. (b) The numerical simulation results for the density of cooperator ($\rho$) at equilibrium for different social population ($N$) at $b$=1.10. 3000 simulations are averaged in each case. }
\end{figure*}
In the investigation above, we choose 50\% as the initial density of cooperators ($\rho_0=0.5$). In this section we would illustrate the effect of different values of $\rho_0$ on the cooperators density at equilibrium ($\rho$). Figure 6(a) shows the real-time fraction of cooperators ($\rho(t)$) for different initial cooperator densities ($\rho_0$). Those data obtained by taking the average value after 3000 times through the same experiment are simulated on the grid of 100x100 at $b$=1.1 and $N$=10000.
As we see, different value of $\rho_0$ can only appreciably affect the the time it takes for the game to reach equilibrium, but do not change the fraction of cooperators at equilibrium. Thus we can say that the fraction of cooperators at equilibrium is insensitive to the initial fraction of cooperators.
\par
Finally, we investigate the effect of the social population ($N$) on cooperation level. We fix $b=1.10$ and $\rho_0=0.5$, then change social population to obtain corresponding fraction of cooperators at equilibrium. The results are showed in Fig.6(b). For $N\ge 800$, $\rho$ does not depend on the social population. However, for $N\le 800$, $\rho$ decreases with smaller $N$. It is clear that a highly cooperative society depends on a sufficient social population.
\section{Conclusions}
We introduce a new dynamic rule of game individual strategy adjustment, that is, monte carlo rule and investigate the prison's dilemma game under it on the grid. Monte carlo rule not only promotes cooperative behavior, but also has higher robustness when compared with unconditional imitation rule, replicator dynamics rule and fermi rule. Under this rule, spatial structure plays a positive role in cooperative behavior, and the equilibrium density of cooperator as a function of
the temptation to defect can be perfectly characterized
by the trigonometric behavior instead of the power-law behavior discovered in the pioneer's work under the fermi rule.
The society obviously welcomes the cooperation: cooperators can obtain higher and stabler returns than defectors throughout the whole temptation parameter ranges.
In addition, the cooperation level is insensitive to the initial density of cooperators but enough social population is needed to maintain a high  cooperation level.



\end{document}